\documentclass[useAMS,usenatbib]{mn2e}
\usepackage{times}
\usepackage{psfig}
\usepackage{graphics}
\usepackage{lscape}
\usepackage{epsf}
\usepackage{epsfig}
\usepackage{color}

\def\changed{}
\title[X-ray emission of stellar clusters]{Evolution of X-ray emission from 
young massive star clusters}
\author[L. M. Oskinova]
{L. M. Oskinova\thanks{E-mail: lida@astro.physik.uni-potsdam.de}\\ 
Astrophysik, Univerit{\" a}t Potsdam, Am Neuen Palais 10,  
Potsdam 14469, Germany}

\begin{document}

\date{Accepted . Received ; in original changedorm }
 
\pagerange{\pageref{firstpage}--\pageref{lastpage}} \pubyear{2005}

\maketitle

\label{firstpage}
    
\begin{abstract}
The evolution of X-ray emission from young massive star clusters is
modeled, taking into account the emission from the stars as well
as from the cluster wind. It is shown that the level and character 
of the soft (0.2-10 keV) X-ray emission change drastically with cluster 
age and are tightly linked with stellar evolution. Using the modern X-ray 
observations of massive stars we show that the correlation between 
bolometric and X-ray luminosity known for single O stars also holds 
for O+O and O+Wolf-Rayet (WR) binaries. The diffuse emission originates 
from the cluster wind heated by the kinetic energy of stellar winds and 
supernova explosions. To model the evolution of the cluster wind, the 
mass and energy yields from a population synthesis are used as input to  
a hydrodynamic model. It is shown that in a very young clusters 
the emission from the cluster wind is low. When the cluster evolves, 
WR stars are formed. Their strong stellar winds power an increasing  
X-ray emission of the cluster wind. Subsequent supernova explosions pump 
the level of diffuse emission even higher. Clusters at this evolutionary 
stage may have no X-ray bright stellar point sources, but a relatively 
high level of diffuse emission. A supernova remnant may become a dominant 
X-ray source, but only for a short time interval of a few thousand years. 
We retrieve and analyse Chandra and XMM-Newton observations of six massive 
star clusters located in the Large Magellanic Cloud. Our model reproduces 
the observed diffuse and point-source emission from these LMC clusters, 
as well as from the Galactic clusters Arches, Quintuplet and NGC\,3603. 
\end{abstract}

\begin{keywords}
stars: winds, outflows -- stars: Wolf-Rayet -- open clusters and 
associations: general -- galaxies: LMC -- X-rays: stars. 
\end{keywords}

\section{Introduction}

Massive star clusters, and super star clusters (SSC) in particular, are
among the most extreme cases of star formation regions in the local
Universe. These clusters have a central stellar density up to $\sim\,10^5
M_\odot\,{\rm pc}^{-3}$ and a total mass ranging from $10^3$ to 
$10^6 M_\odot$.
Clusters of such mass and density can contain hundreds of thousands
of OB stars, providing  a unique laboratory to test the theory of
evolution of massive stars. Moreover, these clusters can have a dramatic
effect on their surrounding interstellar, and in some cases,
intergalactic medium. Among Galactic examples of SSCs are the Arches and
the Quintuplet cluster located in the vicinity of the Galactic center, and
NGC\,3603. A few dozens of massive star clusters are known in
the Magellanic Clouds.

Recent advantages in X-ray imaging and spectroscopy, resulting in the
detection of X-ray emission from some of the SSCs, prompted
theoretical work to explain this emission. It is understood that 
regions of active star formation have high supernova (SN) rates. If 
the supernova energy input is thermalised, a strong wind is driven out 
of the active region. \citet{CC85} presented a self-similar analytical 
solution of the hydrodynamic
equations for the wind driven from a region of uniform mass and energy
distribution. This solution is scalable with the energy and
mass input rates, and with the radius of the region of mass and energy
production. 

% ----------------  Figure 1 ------------------------------------------
\begin{figure*}
\epsfxsize=\textwidth
\centering \mbox{\epsffile{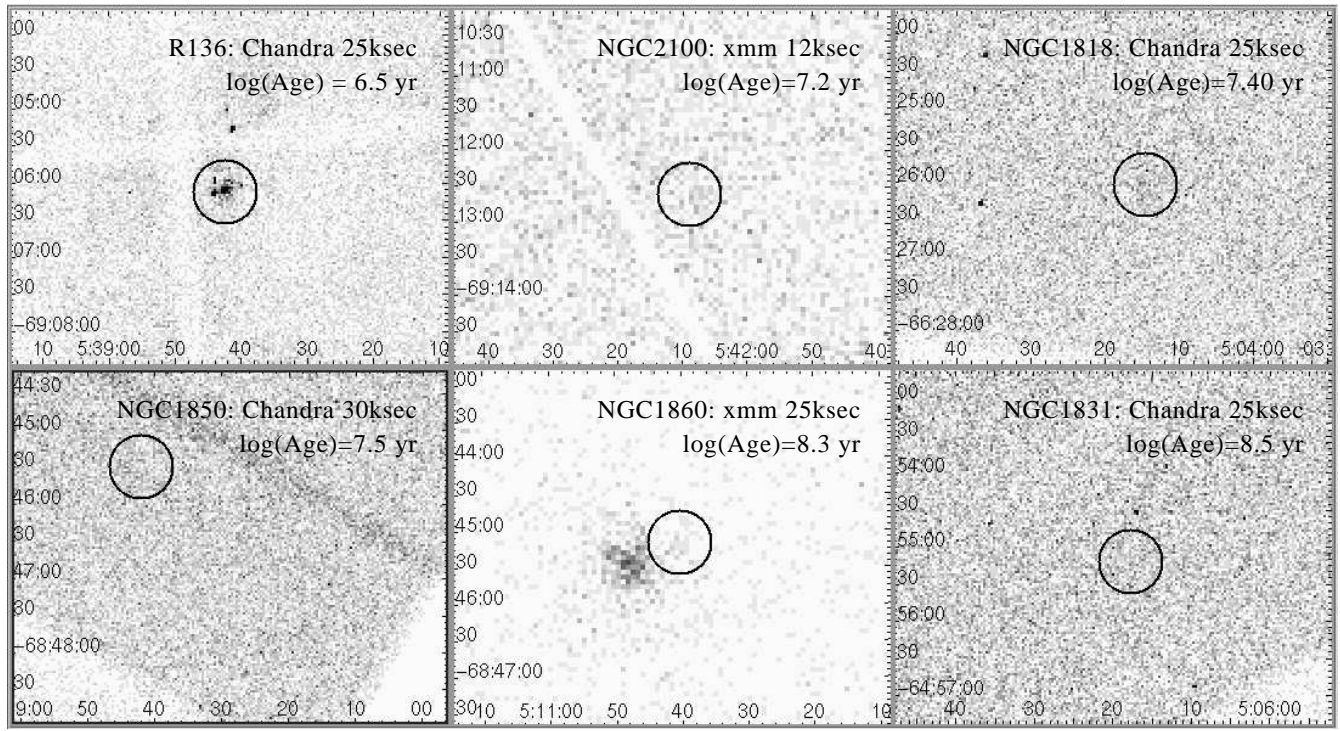}}
\caption{XMM-Newton and Chandra images of star clusters in the LMC.
The celestial positions of the clusters are indicated by arrows. R\,136
is   clearly visible in the top left frame, while the other five
clusters are hardly seen. Note that R\,136, the brightest cluster, 
is by far the youngest 
cluster in the sample. The cluster coordinates and ages are from
\citet{mackey03}}
\label{fig:LMC}
\end{figure*}
%---------------------------------------------------------------------

Canto, Raga \& Rodriguez (2000) considered a ''cluster wind`` resulting 
from the multiple
interactions of stellar winds produced by the massive stars of a dense
cluster. Their cluster wind model is essentially the same as presented
by \citet{CC85}, but accounts for mass loading due to  stellar winds
instead of supernovae.  An analytical solution from of the 
hydrodynamic equations was obtained. A numerical simulation of a
cluster of 30\,stars was also presented. The authors applied their
result to the Arches cluster and predicted that the X-ray emission from
its cluster wind can be detectable. \citet{law04} have reported the
discovery of diffuse X-ray emission from the Arches on a level similar
as calculated by \citet{canto}, confirming their prediction.

%========================================================================
\begin{table*}
\caption{X-ray observations of LMC massive star clusters}
\smallskip
\begin{center}

\begin{tabular}{cccccc}
\hline
\noalign{\smallskip}
Cluster & \multicolumn{2}{c}{Center (J2000.0)}& Detector & Exp. time & Count Rate \\ 
name    & $\alpha$ & $\delta$  & name     & [ksec]   & [ct\,s$^{-1}$]   \\  \hline
\noalign{\smallskip}
R\,136    & 05:38:42.5 & -69:06:03 &ACIS-I  & 21        &$8.31\times 10^{-2}$ \\
NGC\,2100 & 05:42:08.6 & -69:12:44 &MOS    & 6.5        &
$6\arcmin$\,{\small off-axis,\,non-det} \\
NGC\,1818 & 05:04:13.8&-66:26:02 & ACIS-S  & 25        &$4.36\times 10^{-3}$ \\
NGC\,1850 & 05:08:41.2&-68:45:31 & ACIS-S  & 30        & {\small on-axis,\,non-det}\\
NGC\,1860 & 05:10:38.9&-68:45:12 & MOS1    & 4.4       & {\small off-axis},\,$5.9\times 10^{-3}$ \\
NGC\,1831 & 05:06:17.4&-64:55:11 & ACIS-S  & 25        & {\small on-axis,\,non-det}\\
\hline 
\multicolumn{6}{l}{\small The coordinates of the cluster centers are from 
\citet{mackey03}}\\
\end{tabular}
\label{tab:lmcobs}
\end{center}
\end{table*} 
%==========================================================================

The evolution of the X-ray emission from a cluster of single young stars
was investigated by Cervi{\~n}o, Mas-Hesse \& Kunth (2002). They proposed  
an evolutionary
synthesis model, where both supernova remnants and hot diffuse gas
contribute to the X-ray emission. However, the hydrodynamics of the hot
diffuse gas was not considered, and it was assumed that the X-ray
luminosity of this gas is just some arbitrary fraction of the kinetic
energy rate released in the cluster.

A theoretical model which includes the effect of mass loading based on 
the work of \citet{CC85} was considered by \citet{Stev03}. The authors
compared the recent X-ray observations of local massive star clusters
(Rosette, NGC346, NGC3603, R\,136 and the Arches) with the predictions
from the cluster wind theory. It was pointed out that from the
observational side the problems remain as to being sure that the
diffuse emission is genuinely diffuse and associated with the cluster
wind. \citet{Stev03} showed that the diffuse X-ray luminosity of a
cluster wind is correlated with the cluster wind kinetic energy divided
by the cluster radius, but the X-ray temperature inferred from the
available data is not well correlated with the predicted one.
The authors concluded that from the available data it remains 
very unclear as to what is going on in the clusters regarding the 
X-ray emission. 

The research we are presenting here is stimulated by the X-ray images of
the Galactic Center obtained with the XMM-Newton and Chandra
observatories. Two rich massive star clusters, the Arches and the
Quintuplet, have comparable masses, are located at approximately the
same distance, suffer similar interstellar absorption, but their X-ray
images, albeit being in the same field of view, are strikingly
different. The Arches exhibits bright point sources and diffuse
emission, while the Quintuplet is barely detectable. Only careful
examination of the Quintuplet by
\citet{law04} revealed the presence of weak X-rays from this cluster.
There are hardly any point sources seen in the Quintuplet, and the ratio
of diffuse X-ray to the IR flux is much higher in the Quintuplet than in
the Arches. The X-rays from the Arches were modeled in a number of
papers (see above), but the scarcity of X-rays from the somewhat older 
and less compact Quintuplet is not explained.

Another sample of star clusters with similar structural parameters, but
different age is known in the Large Magellanic Cloud (LMC)
\citep{mackey03}. Conveniently, we could use the archival X-ray
observations of six massive LMC clusters. Quite similar to the Galactic
Center clusters, the youngest and most compact cluster R\,136 is bright
in X-rays, while older clusters appear as non-detections, at least from
the first glance.

In the present paper we attempt to explain these observations as a
consequence of cluster evolution. For our study we chose an approach
similar to \citet{Stev03}, but make use of the self-similar properties
of the cluster wind solution.  The effects of stellar evolution on the
mass and energy deposition in a cluster are included via an
evolutionary synthesis model. The important novelty is that we consider
stellar X-rays as well as diffuse emission. It is often impossible to
disentangle the diffuse and stellar emission in the dense cluster core.
Therefore we estimate the luminosity of the stellar population and add
it to the predicted luminosity of the cluster wind. The total, then,
can be  compared with observations. The stellar X-ray emission changes
significantly when a star evolves, and we account for these changes as
well. Our analysis is restricted to the inner parts of the clusters where
the bulk of mass and energy is deposited. This allows us to concentrate
on the properties of cluster winds and stellar population, and to avoid
considering the hot bubbles that are expected around regions with
fast winds.

The paper is organised as follows. The observations of massive stellar
clusters used in this work are described in  Section\,2. The approach to
the modeling is specified in Section\,3. The evolution of X-ray emission
of cluster wind is calculated in Section\,4. Section\,5 is devoted to
the evolution of the stellar X-ray luminosity. In this section we pay
special attention to the correlations between X-ray and bolometric
luminosities. In Section\,6 we discuss the X-ray emission of supernova
remnants, and conclusions are drawn in Section\,7.

\section{Observations of young massive  clusters}

% ----------------  Figure 2 ------------------------------------------
\begin{figure*}
\epsfxsize=13cm
\centering \mbox{\epsffile{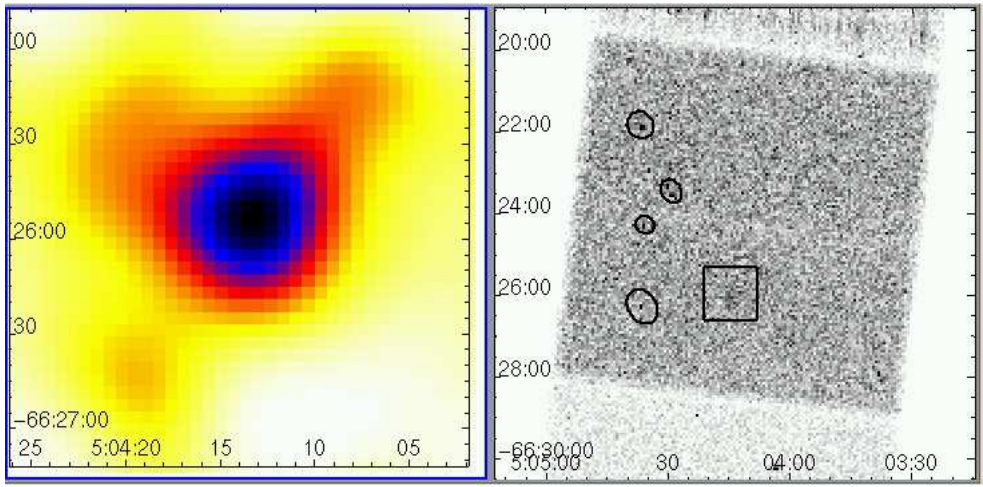}}
\caption{ {\it Right panel}: An ACIS-S image of the region of NGC\,1818.
produced from the event list binned by the factor of 8. 
The standard software detects NGC\,1818 (within the square box), as well as 
other sources (encircled). {\it Left panel:} Enlarged area within the square 
box in the right panel with the adaptively smoothed image of NGC\,1818.}
\label{fig:ngc1818}
\end{figure*}  
%----------------------------------------------------------------------------

Recently a sample of six massive young star clusters located in the 
LMC was observed with the XMM-Newton and Chandra X-ray observatories. The 
observations are in the public archives (see Table\,\ref{tab:lmcobs}). 
{\changed To our knowledge the results of these observations, except 
of R\,136,  
have not been published. Therefore we retrieved the data on five LMC 
clusters from archives and analysed them.} All clusters have similar masses, 
but differ significantly in age. The structural parameters of the clusters 
are taken from \citep{mackey03} and listed in Table\,\ref{tab:lmc}. 
Figure\,{\ref{fig:LMC}} shows the X-ray images of the areas of the sky, 
where the clusters are located. Although being at the same distance and 
observed through similar absorbing columns, the clusters X-ray images are 
strikingly different.

The youngest and most compact cluster, R\,136, is at least one order of
magnitude brighter than the more evolved clusters and displays wealth of
bright point sources, but little  diffuse emission. Among more
evolutionary advanced clusters, three are not detected and the detected
ones are apparently lacking point sources. 

In order to estimate observed X-ray fluxes  from these LMC clusters, a
standard analysis  of pipeline processed  products has been performed.
Retrieved count rates were transformed  to fluxes  assuming that
radiation comes from thermal bremsstrahlung with  $kT_{\rm X}=3$\,keV.
This assumption is based on the spectral fits of the diffuse emission
from the  Arches and the Quintuplet clusters as discussed  in
\citet{law04}.The neutral hydrogen column density for all LMC clusters
was adopted as $N_{\rm H}=5\times 10^{21}$\,cm$^{-2}$, as derived from  
spectral fits of R\,136 stars by Portegies Zwart, Pooley, \& Lewin (2002). 

Below we briefly describe how the diffuse X-ray luminosity was estimated 
for  each cluster under consideration. 

R\,136 was observed by ACIS-I on board of the Chandra X-ray observatory  
for 21\,ksec exposure time. However, even with the superb angular resolution 
of the  ACIS camera (the pixel size is $0.49\arcsec$), it is hardly
possible to separate stellar and diffuse emission within the cluster core,
which has a radius $1.32\arcsec$ (corresponding  to 0.32\,pc).
However, as \citet{mackey03} pointed out from the analysis  of the
cluster brightness profile, the inner region of the cluster is 
extended to 2.43\,pc. We estimate the diffuse X-ray luminosity of R\,136
from this inner region. 

Using the {\it wavedetect} tool for the binned by factor of 6 event
file we obtained the  total count rate in this region and translated it
into the background subtracted flux,  $F_{\rm X}=1.6\times
10^{-12}$\,erg\,s$^{-1}$\,cm\,$^{-2}$. Background correction is
important  since R\,136 is embedded in the extended region of diffuse
X-ray emission 30\,Dor\,C.  As the next step, we subtracted the
luminosities of the point sources CX1$..$9 \citep{zwart02}, and
attributed the remaining luminosity to the diffuse emission. 

NGC\,2100 was observed $6\arcmin$ off-axis by XMM-Newton.
{\changed Although clearly seen on an OM image}, this rich young cluster
is not detected in the X-rays  (see Fig.\,\ref{fig:LMC}). An
upper limit to the count rate from an area corresponding to the cluster 
core was obtained from a sensitivity map, and
is   $7\times 10^{-3}$\,ct\,s$^{-1}$ for the pn\,camera. Thus the upper 
limit on the flux from NGC\,2100 in the XMM-Newton  passband is 
$F_{\rm X}< 3.5\times 10^{-14}$\,erg\,s$^{-1}$\,cm\,$^{-2}$. 

%================================ Table 2 =================================
\begin{table*}
\caption{Parameters of the massive star clusters}
\smallskip
\begin{center}

\begin{tabular}{ccccccccr}
\hline
\noalign{\smallskip}
Cluster&\multicolumn{2}{c}{Center (J2000.0)}& $r_c$ & $r_c$ &log Age & Metallicity &
$\log M_{\rm cl}$ &$\log {L_{\rm X}^{\rm diff}}$ \\
name  &  $\alpha$ & $\delta$ & [arcsec]   & [pc]  & [yr]  & [Fe/H] & 
$[M_\odot]$ & [erg\,s$^{-1}$] \\ \hline
\noalign{\smallskip}
R\,136    & 05:38:42.5&-69:06:03 & 10   & 2.43 & 6.5 & $\sim -0.4$ & $4.40$ & 34.3\\
NGC\,2100 & 05:42:08.6&-69:12:44 & 5.02 & 1.22 & 7.2 & -0.32 & 4.31& $< 34.04$  \\ 
NGC\,1818 & 05:04:13.8&-66:26:02 & 10.10& 2.45 & 7.4 & 0.0\,to\,-0.4&$4.01$ &34.32 \\
NGC\,1850 & 05:08:41.2&-68:45:31 & 10.48& 2.55 & 7.5 & -0.12 & 4.87 &$< 32.70$  \\
NGC\,1860 & 05:10:38.9&-68:45:12 & 9.00 & 2.19 &8.3  & -0.52 & 3.90 &34.61     \\
NGC\,1831 & 05:06:17.4&-64:55:11 & 18.28& 4.44 &8.5  &+0.01  & 4.71 &$< 33.48$\\  
NGC\,3603 &11:15:09.1 &-61:16:17 & 29.5 & 1.00 & 6.0 & 0.02 & 4.0 &34.3$^1$ \\
Arches    &17:45:50.4 &-28:49:20 & 10   & 0.39 & 6.3 & 0.02 & 4.9 &34.2$^2$ \\   
Quintuplet&17:46:14.8 &-28:49:35 & 12   & 0.46 & 6.6 & 0.02 & 3.8 &33.3$^2$ \\  
\hline 
\noalign{\smallskip}
\multicolumn{9}{l}{\small Structural parameters of LMC clusters are from 
\citet{mackey03}}\\
\multicolumn{9}{l}{\small $^1$ from \citet{sung} and \citet{ngc3603}}\\
\multicolumn{9}{l}{\small $^2$ from \citet{arfig}, 
Figer, McLean \& Morris (1999) and \citet{law04}}    
\end{tabular}
\label{tab:lmc}
\end{center}
\end{table*} 

%============================================================================

The Chandra X-ray observatory has performed 25 ksec pointing observations
of NGC\,1818 and NGC\,1831, and a 30 ksec pointing observation of NGC\,1850. 
NGC\,1818 is a faint source of apparently diffuse emission detected with the 
standard Chandra data analysis  {\it wavedetect} tool. 
{\changed The estimated number of source counts is 109. They are distributed nearly 
uniformly over a $15\arcsec\times 15\arcsec$ area of the sky, with the center 
of diffuse emission coinciding with the cluster center as listed in the 
Table\,\ref{tab:lmc}. Additionally, the {\it psfratio} determined 
by {\it wavedetect} is $>1$, indicating that the source indeed may 
be extended. The low signal-to-noise spectrum of NGC\,1818 was extracted.

In order to gain information on the absorption column density we
extracted and fitted the spectrum of the brightest background object
present in the ACIS-S field of view. The Seyfert\,I galaxy Cal~F, only
$10\arcmin$ away from NGC\,1818, is located behind the LMC \citep{geha}.
We tentatively fitted the spectrum of Cal~F with a two-component model
(power law and black body). This spectral fit allows to determine the
absorption column density in the direction to NGC\,1818, $N_{\rm
H}\approx 1.4\times 10^{21}$\,cm$^{-2}$. One can make the plausible
assumption that the absorption is mainly interstellar and is not
intrinsic to the Seyfert\,I AGN.

With the absorption column density constrained, the spectrum of the X-ray 
emission from NGC\,1818 can be fitted either with a power law with index 
$\approx 2.5$, or with a multi-temperature thermal plasma with temperatures 
ranging from 0.3 to 5 keV.} The raw and adaptively smoothed images of the 
cluster are shown in Fig.\,\ref{fig:ngc1818}. The X-ray flux from NGC\,1818 
is $F_{\rm X}=7\times 10^{-14}$\,erg\,s$^{-1}$\,cm\,$^{-2}$.  

A similar observation of NGC\,1831 and NGC\,1850 failed to detect X-rays 
from these cluster. In order to estimate an upper limit on its X-ray flux 
we run the {\it wavedetect} tool with  significantly lower detection 
threshold, therefore allowing for fake detections. Eventually the area which 
coincides with the location of the clusters  was ``detected``,  providing an 
estimate of the count rate. The upper limit on the flux from NGC\,1831 is 
$F_{\rm X}<4\times 10^{-14}$\,erg\,s$^{-1}$\,cm\,$^{-2}$, and from NGC\,1850 
$F_{\rm X}< 10^{-14}$\,erg\,s$^{-1}$\,cm\,$^{-2}$. 
    
NGC\,1860 was observed $9\arcmin$ off-axis by XMM-Newton, and is included 
in the XMM-Newton Serendipitous Source  Catalog. The flux retrieved from the
catalog is  $F_{\rm X}=(7.75\pm 1.8) \times
10^{-14}$\,erg\,s$^{-1}$\,cm\,$^{-2}$. NGC\,1860 is located close to
the X-ray source 1AXG\,J051054-6844   (see Fig.\,\ref{fig:LMC}). The
latter is identified as a foreground late-type star  in the
literature. 

Based on the distance of the LMC, which we adopt as 50.1\,kpc throughout  
this paper \citep{mackey03}, the X-ray fluxes corrected for 
the interstellar absorption are converted into luminosities
(Table\,\ref{tab:lmc}). 

The most massive Galactic star clusters also have been observed in X-rays, 
and analysis of their X-ray properties is available in the literature 
(Table\,\ref{tab:lmc}). Galactic clusters demonstrate the same trend as 
LMC clusters, with younger ones being more X-ray active.  NGC\,3603 
($1\pm 1$ Myr old) exhibits a wealth of stellar point sources, along with 
a detectable level of diffuse emission \citep{ngc3603}. The two most 
massive known  in the Galaxy are the Arches (1-3 Myr old) and the 
Quintuplet (3-5 Myr old) \citep{qufig,arfig}, located in the central 
50\,pc of the Milky Way. \citet{law04} pointed out that the character 
of X-ray emission  of the Arches and Quintuplet is different. The ratio 
of masses as well as of IR luminosities  between Arches and Quintuplet 
is roughly 3:1. However, the ratio of X-ray flux (0.5-8\,keV) is 
approximately 11:1; if only the point source emission is considered, 
the ratio rises  to 18:1. The Quintuplet cluster appears to have relatively 
more diffuse emission, the hottest gas being located in the cluster core. 
In contrast to the Arches' diffuse emission, the Quintuplet's diffuse 
emission is clearly thermal.  

\section{Modeling of X-ray emission from star clusters}

We consider an idealised star cluster with mass $M_{\rm cl}=10^6
M_{\odot}$. It is assumed that the stars are coeval, i.e. the cluster
has undergone an instantaneous  burst of star formation. We assume that
the stellar masses are distributed according to the standard 
(Salpeter) initial mass function (IMF), $\xi(M)=M_0 M^{-2.35}$, with a
lower mass cut-off at $M_{\rm min}=1\, M_\odot$ and an upper mass cut-off 
at  $M_{\rm max}=100\, M_\odot$. The radius of the cluster is
such that the central density is  about $10^5\,M_\odot\,{\rm pc}^{-3}$.
The calculations are done for solar ($Z=0.02$) and LMC ($Z=0.008$) 
metallicity. 

With the above assumptions, the output from the \citet{sb99} stellar
population synthesis  code can be directly applied.   

The major simplifications we have made are the following: (1) We
effectively assume a  cluster of single stars. The realistic account
for binary stars in the  cluster would affect the IMF and the age
attributed to the cluster. We account for the
binarity in a parameterised way when modeling the cluster wind.
We also consider binaries in estimates of the collective stellar X-ray
luminosity. (2) We neglect  the cluster dynamics. \citet{mackey03}
analysed surface brightness profiles and structural parameters for
53 clusters in the LMC. They concluded that the spread in the core
radius  increases significantly with increasing cluster age. For our
model cluster  the  radius does not change during the evolution.
Cluster mass is also  assumed to be constant. (3) We assume that the
mass-luminosity relation  holds for the stars of all masses. We also
assume that the stars evolve  with constant mass till they reach
WR phase.  (4) We do not consider X-ray emission from degenerate  stars
and X-ray binaries. There is a broad discussion in the literature on
this subject. (5) We exclude any possible sources of non-thermal X-rays.
       
The time dependent X-ray luminosity of a cluster is 
\begin{equation}
   L_{\rm X}(t)\,=\,L_{\rm X}^{\rm point}(t)\,+\,L_{\rm X}^{\rm
   diff}(t).
\end{equation}
The diffuse X-ray emission, $L_{\rm X}^{\rm
diff}$, originates in the cluster wind. The collective stellar 
luminosity is
\begin{equation}
L_{\rm X}^{\rm point}(t)=\,\int_{M_{\rm min}}^{M_{\rm max}} \xi(M)\
L_{\rm X}(M,t)\ dM,
\label{eq:lxpoi}
\end{equation}
\noindent where \(L_{\rm X}(M,t)\) is the X-ray luminosity of a star with
mass $M$  at age $t$. In the next two sections we describe our modeling
of the diffuse and the stellar component.

\section{Diffuse X-ray emission of star clusters}

\subsection{Wind from a model cluster} 

The interstellar medium (ISM) in a star cluster receives a constant
supply of gas, at a rate $\dot{M_\ast}$, due to the mass loss from 
massive stars and, later on, supernova  explosions. In addition,
material can be ablated from protostellar (or planetary) disks 
\citep[e.g.][]{kroupa}. Some quantities of primordial gas may also be 
present in a young cluster. Following \citet{Stev03}, let us call this 
additional contribution $\dot{M}_{\rm cool}$. Then the total mass input
to a  star cluster is $\dot{M}=\dot{M_\ast}+\dot{M}_{\rm cool}$.  Stellar
winds and SN explosions  supply also kinetic energy to the cluster's
ISM at a  rate $\dot{E}$. Some fraction of this kinetic energy input is
thermalised, heating the gas to X-ray  emitting temperatures.  Both,
$\dot{M}$ and $\dot{E}$, vary significantly during evolution of the
cluster.  Being initially low for a very young cluster, stellar
mass-loss and energy input  are increasing while stars are evolving, and
reach their maximum with the beginning  of SN ignition. On the other
hand, in an old cluster, where massive stars are absent,  $\dot{M}$ and
$\dot{E}$ steeply decline. 

% ----------------  Figure 3 ------------------------------------------
\begin{figure}
\epsfxsize=\columnwidth
\centering \mbox{\epsffile{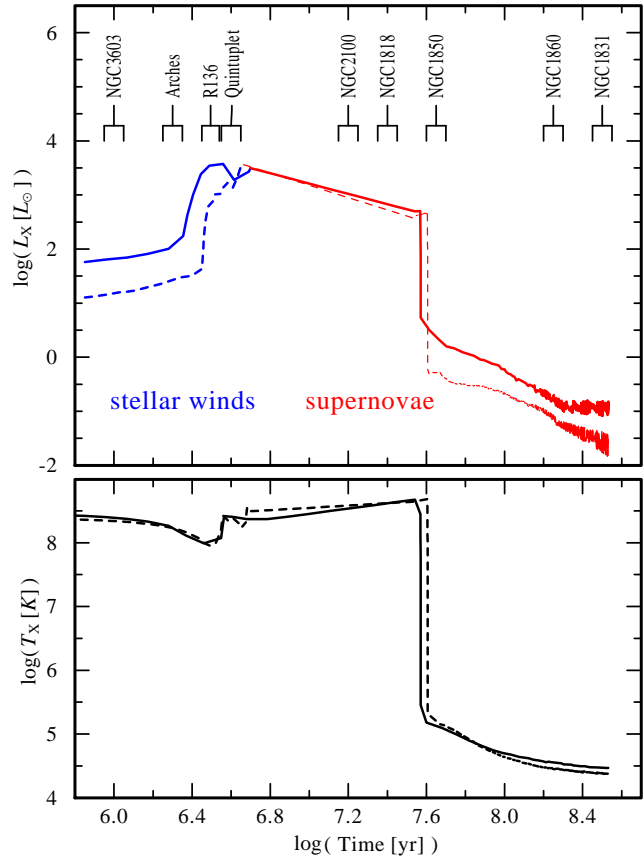}}
\caption{X-rays from a model cluster of $10^6\,M_\odot$. 
{\it Upper panel}: Luminosity of cluster wind versus
cluster age for solar metallicity (solid line)
and LMC metallicity (dashed line). The epochs when mass and energy 
input is dominated a) by stellar  winds and b) by supernova explosions
are indicated. The part of the curves between 5 and 35 Myr is artificially 
smoothed to correct for the computational discontinuity of Starburst99 
models  for SN rates \citep[for details see][]{sb99}. {\it Lower panel}: 
Temperature of the cluster wind as function of the cluster age. Solid and 
dashed lines are for the solar and LMC models, respectively. Note 
that after $\approx\,40$ Myr of cluster evolution, the temperature becomes 
too low for X-ray production.}  
\label{fig:lxdif}
\end{figure}  
%-----------------------------------------------------------------------------

% ----------------  Figure 4 ------------------------------------------
\begin{figure}
\epsfxsize=\columnwidth
\centering \mbox{\epsffile{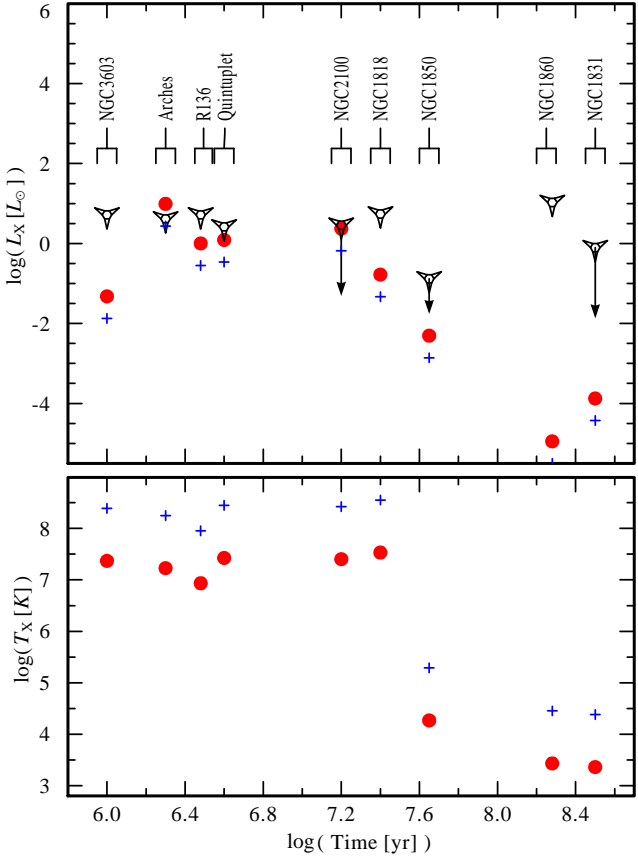}}
\caption{{\it Upper panel}: Comparison of the observed diffuse X-ray
luminosities (diamonds) with individual cluster models. Crosses are 
the theoretical diffuse X-ray luminosity assuming that $\alpha=1$ and 
$\dot{M}_{\rm cool}=0$  (see text).  Circles are the model values
assuming $\alpha=0.1$ and  $\dot{M}_{\rm cool}=0.05\dot{M_\ast}$.
{\it Lower panel}: Temperature of the cluster wind for a model with 
$\alpha=1$ and $\dot{M}_{\rm cool}=0$ (crosses) and 
$\alpha=0.1$ and $\dot{M}_{\rm cool}=0.05\dot{M_\ast}$ (circles).}  
\label{fig:obscl}
\end{figure}  
%-----------------------------------------------------------------------------

Stellar population synthesis, such as  Starburst99 \citep{sb99}, are
convenient tools to provide $\dot{M}$  and $\dot{E}$ that account for
both, stellar winds and supernova explosions. We used here the
standard output  of Starburst99 for the appropriate metallicities to
obtain the mass and energy yields. 

There is a shortcoming in Starburst99 for determining the mass and energy 
input in a dense star cluster.  The model accounts only for single stars,
while in fact the majority of high-mass stars in massive clusters are in 
binary or multiple systems. The mass-loss rate from a binary star is 
basically the sum of the mass-loss rates from two single stars. However, 
there can be a difference in the rates of kinetic energy
input between  single and binary stars. In massive binary systems the
stellar winds interact in a  way, that a part of the total kinetic
energy is dissipated,  e.g.\ for the heating of the colliding winds
zone, and radiated away. Although it is   possible to speculate which
fraction of kinetic energy is consumed due to the
wind-wind collision, there are a number of other uncertainties regarding 
energy losses in a star cluster. A parameter $\alpha$, accounting for all
these  uncertainties, was introduced in \citet{CC85} and later called
``thermalisation   efficiency'' in \citet{Stev03}. Thus, the energy 
losses due  to the wind-wind collisions in binary systems can be
accounted  for in a parameterised way, using factor $\alpha$.  

To calculate the X-ray luminosity of cluster wind  one must solve the 
hydrodynamical equations. \citet{CC85} have shown that a wind from a star 
formation region can be described as a self-similar process \citep{ZR67}, 
and obtained solutions  applicable to mass loaded cluster winds.

Let us consider the evolution of the broad band X-ray luminosity of an
idealised model cluster {\changed with mass $10^6\,M_\odot$ and radius 
$R_{\rm cl}=2$\,pc. The resulting density is $3\times 10^4 M_\odot\,{\rm
pc}^{-3}$.} We consider the X-ray luminosity of the whole cluster core, 
and do not infer its radial profile. Therefore the \citet{CC85}
solutions for density, $\rho(r)$, and the pressure, $P(r)$,  are
integrated over the cluster core radius.

The frequency-integrated  X-ray luminosity of the cluster wind at epoch 
$t$ is 
\begin{equation}
   L_{\rm X}^{\rm diff}(t)\,=\,\Lambda(T_{\rm X})\ E\!M(t),
\label{eq:Ldif}
\end{equation}

\noindent where $\Lambda(T_{\rm X})$ is the cooling function, and $
EM(t)$  is the emission measure of the cluster wind:

\begin{equation}
  E\!M(t)\,=\,\int_{\rm V}\frac{\rho(t)^2}
{\mu_{\rm i}\mu_{\rm e}m_{\rm H}^2}\ dV,
\end{equation} 

\noindent and

\begin{equation}
   T_{\rm X}(t)\,=\,\frac{m_{\rm H}\mu_{\rm i}}{k}\frac{P(t)}{\rho(t)},
\end{equation}
with the Boltzman constant $k$ and proton mass $m_{\rm H}$. The mean 
molecular weights per ion, 
$\mu_{\rm i}$ and per electron, $\mu_{\rm e}$, depend on the cluster 
age $t$ due to chemical evolution. However, for simplicity we assume 
that  $\mu_{\rm i}$ and $\mu_{\rm e}$ do not change significantly and 
consider them to be constant ({\changed 1.3 and 1.14, respectively}).

From scaling laws of self-similar solutions, emission measure and
temperature of the  cluster wind within the mass and energy production
region of radius $R$ are

\begin{equation}
E\!M(t)\,=\,1.4\times 10^{-2}\,\frac{4\pi}{\mu_{\rm i}\mu_{\rm e}m_{\rm H}^2}\,
\frac{\dot{M}^3}{R\alpha\dot{E}},
\label{eq:em}
\end{equation}

\noindent and

\begin{equation}
T_{\rm X}(t)\,=\,0.4\,\frac{m_{\rm H}\mu_{\rm i}}{k}\,\frac{\alpha\dot{E}}{\dot{M}}.
\label{eq:tx}
\end{equation}

\noindent Therefore, when $\dot{M}$ and  $\dot{E}$ are known at time
$t$, the temperature, density and velocity of  the cluster wind can be
obtained without solving partial  differential equations.

Assuming  $\alpha=1$ and $\dot{M}_{\rm cool}=0$, we substitute the mass
input, $\dot{M}(t)$, and kinetic energy input, $\dot{E}(t)$,  provided
by population synthesis  into Eqs.\,(\ref{eq:Ldif}, \ref{eq:em}, 
\ref{eq:tx}) and obtain the luminosity of the cluster wind versus time
(Fig.\,{\ref{fig:lxdif}}).

As can be readily seen from Fig.\,{\ref{fig:lxdif}}, the level  of
diffuse emission varies significantly during the cluster history.  In a
cluster younger than 2.4\,Myr ($\log {\rm Age}=6.4$), stellar 
population synthesis predicts neither stars with high mass-loss rates 
(such as LBVs or WR stars) nor supernova remnants. The influx of mass
into the intracluster medium is therefore small, resulting in a 
relatively low density and emission measure. 

After the first $\approx\,2.4$\,Myr of stellar evolution the most
massive stars pass trough the  LBV stage, which is a very brief but
nevertheless important phase because of high mass loss. Subsequently
they evolve to Wolf-Rayet stars. Due to the mass and energy input
of these stars, the level of diffuse X-ray emission of the cluster wind
increases by two orders of magnitude. WR stars finish their lives after
a few\,$\times 10^5$ years by a SN explosion.  6.3\,Myr
($\log {\rm Age}=6.8$) after initial starburst the  mass and energy
input from SN explosions becomes dominant and maintains a nearly 
constant X-ray luminosity of the cluster wind for millions of years. 

It is not expected that stars with initial masses below $8\,M_{\odot}$
undergo a SN  explosion (Woosley, Heger \& Weaver 2002). These stars  
do not have significant stellar winds either. Therefore, in a cluster 
older then $\sim 40$\,Myr ($\log {\rm Age}=7.6$) no stars with high mass 
loss are left. Consequently the mass and energy supply to the cluster wind 
is cut-off. The temperature drops  sharply and cluster wind is not a
source of diffuse X-ray emission any more.  

\subsection{Comparison with observations}

{\changed
The above theoretical considerations can now be confronted with the
empirical data described in Section\,3. First, we have plotted the 
observed diffuse X-ray luminosity of the clusters listed in 
Table\,{\ref{tab:lmc}} 
in the upper panel of Fig.\,{\ref{fig:obscl}} (diamond symbols).
Next, for each  cluster we calculated the theoretical X-ray luminosity and
the temperature of cluster wind, assuming  $\alpha=1$, and 
$\dot{M}_{\rm cool}=0$. 
The crosses in Fig.\,{\ref{fig:obscl}} represent these theoretical values.
As can be seen from the upper panel of Fig.\,{\ref{fig:obscl}} the observed 
diffuse X-ray luminosity is much higher than the theoretical one, the only 
exception being the Arches.

Given the quality of the data, the information on the observed temperature 
is quite limited. \citet{law04} analysed the spectra of the diffuse gas in 
the Arches and in the Quintuplet. For the Arches, the best fit is a 
two-temperature thermal model with $T_1=1.4^{+0.5}_{-0.9}\times 10^7$\,K 
and $T_2=7.2_{-6.6}\times 10^7$\,K. For the Quintuplet the peak temperature 
of the diffuse gas in the cluster center is $T=(3\pm 0.6) \times 10^7$\,K. 
\citet{Stev03} report $2\times 10^7$\,K as the temperature of diffuse gas 
in R\,136. Our analysis of NGC\,1818 suggests that the observed temperature 
is ranging from $4\times 10^6$~to~$6\times 10^7$\,K. (However, due to the 
quality 
of the spectra the question whether the Arches and NGC\,1818 emission is thermal 
remains open.)  As can be seen from the lower panel in Fig.\,{\ref{fig:obscl}}
the theoretical X-ray temperatures (crosses) are apparently too high 
compared to the temperature estimates deduced from the spectral fits.

There are two aspects to be considered in order to explain these discrepancies.} 
First, the luminosities of diffuse  emission listed in Table\,\ref{tab:lmc} 
cannot be attributed unambiguously to the cluster wind alone. Second, regarding 
the model, we adopted a certain cluster age (listed in  Table\,\ref{tab:lmc}). 
The estimates of the cluster ages, however, have a large uncertainty. E.g., the 
estimated ages are  1\,--\,3\,Myr for the Arches, 3\,--5\,Myr for the 
Quintuplet, 
and  $1\pm 1$\,Myr for NGC\,3603. As can be seen from Fig.\,{\ref{fig:lxdif}}, 
the model predicts that the diffuse X-ray luminosity rises by two orders of 
magnitude between 1.6\,and\,3.2\,Myr. Hence, the predicted X-ray luminosity 
is very sensitive to the cluster age. 

Second, the values represented by small crosses are obtained under
the assumption  that the mass-loss is only from stellar winds and SNe
($\dot{M}_{\rm cool}=0$) and  that all mechanical luminosity released
in the cluster is spent to heat the cluster wind ($\alpha=1$). But, as
can be seen from Eq.\,(\ref{eq:tx}), a thermalisation efficiency
$\alpha$ smaller then unity, and increased mass input will act to
decrease the temperature of the cluster wind. As pointed out by
\citet{Stev03}, this decrease of the temperature can be understood as a
drop in   the average energy available per particle. Since
bremsstrahlung scales with the square root of temperature and using
scaling relation Eq.\,(\ref{eq:tx}), one notices  that  $L_{\rm X}\propto
\dot{M}^3({\dot{M}}{\dot{E}})^{-1/2}$. Therefore,  for somewhat higher
mass input and reduced energy input, the luminosity is expected to
rise.  

We did not attempt to infer $\dot{M}_{\rm cool}$  and $\alpha$ from the
observations because of the many uncertainties in observed quantities. 
But it can be  shown that it is possible to find such a combination of
these parameters  that  would make the theoretical values of X-ray
luminosity and temperature resembling the observed ones.  As an
example, we plotted in Fig.\,\ref{fig:obscl} (large dots) $L_{\rm
X}^{\rm diff}$ and $T_{\rm X}$ that were obtained assuming  plausible
values $\dot{M}_{\rm cool}=0.05\dot{M}_\ast$ and thermalisation
efficiency  $\alpha=0.1$. Compared to the models with  $\dot{M}_{\rm
cool}=0$ and  $\alpha=1$ (crosses) the luminosity is higher while the
temperature is lower. 

The theoretical values fit quite well to the observed ones for the R\,136
and the  Quintuplet. For the Arches, the model predicts
higher diffuse  X-ray luminosity than is actually observed. There
are at least two explanations for this discrepancy. First, if
the age of the Arches is actually smaller  than assumed here (2\,Myr), 
the predicted X-ray luminosity will be reduced  towards the observed
one. Second, the diffuse luminosity of Arches may 
be higher than the one used here ($1.6\times 10^{34}$\,erg s$^{-1}$),
depending on the interpretation of the observational data. For
instance, \citet{Stev03} used $L_{\rm X}^{\rm diff}=5\times 10^{35}$
erg s$^{-1}$ as the empirical value. 

Two LMC clusters, NGC\,1818 and NGC\,1860, show X-ray emission at a
much higher level than expected from the cluster wind alone. To
explain this, other sources of  X-rays radiation should be invoked. From
the image of NGC\,1818 in Fig.\,\ref{fig:ngc1818} it is  difficult to 
decide if the source is actually diffuse or pointlike. Also,   
the very low count-rate does not allow to extract any useful spectral
information. However, one may expect the presence of a supernova  remnant 
in a 25 Myr old star cluster like NGC\,1818. In fact our sample 
includes five LMC clusters older than 4\,Myr. Statistically, at least 
one of them should exhibit a young supernova remnant (SNR). Applying 
scaling relations for the dynamical age of SNR and SN energy from 
Hughes, Hayashi \& Koyama (1998) to NGC\,1818, we estimate the possible 
SNR age of a few$\times 10^3$\,years and SN's energy of about 
$\approx\,10^{51}$\,erg. The SNR X-ray emission further discussed in 
Sect.\,6.

The relatively high level of X-ray emission from the 200\,Myr old cluster
 deserves special attention. The cluster was observed $9\arcmin$ 
off-axis. It is also $38\arcsec$ apart from a bright foreground star. 
As listed in the XMM-Newton Serendipitous Source Catalogue, NGC\,1860 has 
a detectable level of emission harder than 4.5\,keV. As we will show in 
the subsequent sections, it is not expected that normal stars would be 
significant X-ray sources at the age of NGC\,1818 and NGC\,1860. 
Therefore, X-rays in NGC\,1860 can be possibly attributed to either a SNR 
or a degenerate star. 

Another interesting result is that the observed X-ray emission from 
NGC\,3603 is  much higher than can be accounted for with the cluster 
wind model. As was  suggested by \citet{ngc3603}, the high luminosity 
of diffuse emission in this  cluster may be attributed to unresolved 
low-mass stars. We will investigate the  contribution of the low- and 
high-mass stars in the next sections.  

\section{Evolution of the X-ray luminosity of stars}

If the X-ray luminosity of a star scales with its bolometric luminosity, 
and the latter scales with the stellar mass,  one can estimate 
the total X-ray luminosity of stars with given mass. 

For simplicity we assume that stars evolve with roughly constant
luminosity. In order to model the evolution of stellar X-ray
luminosity of a cluster, we a) calculate the number of stars per mass
bin from the adopted IMF;  b) use a mass-luminosity relation to
estimate the stellar bolometric luminosity;  c) estimate  the X-ray
luminosity using a correlation between X-ray and  bolometric 
luminosities; d) estimate the lifetime in hydrogen-burning stages 
for a star of given mass. Thus a
correlation between X-ray luminosity  and age is obtained.

We have adopted the mass-luminosity relation from \citet{ga},
\begin{equation}
\frac{L_{\rm bol}}{L_{\odot}} =
\left\{ \begin{array}{ll}
81\left(\frac{M}{M_{\odot}}\right)^{2.14} & M > 20 M_{\odot}\\
1.78\left(\frac{M}{M_{\odot}}\right)^{3.5} & 2< M < 20 M_{\odot}\\
0.75\left(\frac{M}{M_{\odot}}\right)^{4.8} & M < 2 M_{\odot},
\end{array} \right.
\label{eq:mlr}
\end{equation}
where $L_{\rm bol}$ is assumed to be equal to the bolometric 
luminosity on the main sequence.

The origin of X-ray emission is different in low and high mass stars.
Low-mass stars ($M_\ast \la 3\,M_\odot$) emit X-rays as result of magnetic 
activity caused by the dynamo mechanism and connected with the presence 
of outer convective layers. Massive stars, on the other hand, are thought 
to be X-ray emitters due to shock waves in their powerful (either
colliding or freely expanding) winds driven by radiation pressure.
Stars with masses between 3 and 10 $M_\odot$ generally do not
emit X-rays, because outer convective zones are absent in
stars more massive than $3\,M_\odot$, and stellar winds are weak 
for stars less massive then $10\,M_\odot$.

% ============================== SECTION 5.1 ====================
\subsection{Low- and solar-mass stars}

%=============================Table 3 =========================
\begin{table}
\caption{Dependence of X-ray luminosity on age for low- and solar-mass
stars \citep[based on][]{flac03}} 
\smallskip
\begin{center}

\begin{tabular}{ccc}
\hline
\noalign{\smallskip}
$M_\ast\,[M_\odot]$ & $0.5\, ...\,1$ & $1\, ...\,3$               \\
\noalign{\smallskip}
$\log$\,(Age)\,[yrs] & $\log\langle L_{\rm X}\rangle$\,[erg/s] &
$\log\langle L_{\rm X}\rangle$\,[erg/s]\\
% years & erg/s & erg/s \\
\hline
\noalign{\smallskip}
5.5 $-$ 6.0 & 30.65 & 31.50 \\
6.0 $-$ 6.5 & 30.0  & 31.18 \\ 
6.5 $-$ 7.5 & 29.51 & 29.73 \\ 
\hline
%\noalign{\smallskip}
%\multicolumn{3}{l}{\small $^1$ based a study of the ONC by \citet{flac03}} \\
\end{tabular}
\label{tab:onc}
\end{center}
\end{table}  
%=======================================================================

To trace the evolution of X-ray emission from an ensemble of low- and 
solar-mass stars ($0.1M_\odot < M_\ast < 3 M_\odot$) we use the results
obtained from the study of stars in the Orion Nebular Cluster (ONC). 
\citet{flac03} explored the relationship between X-ray activity and 
stellar mass and age for 696 well-characterized ONC members based on 
Chandra observations. They found a direct correlation between $L_{\rm X}$ 
and mass for stars less massive than $3\,M_\odot$. For stars with masses 
in the range from 3 to 10\,$M_\odot$, $L_{\rm X}$ is small and attributed 
to the less massive companion in binaries. For massive O stars 
($M > 10\,M_\odot$), $L_{\rm X}$ rises again \citep[see][Fig.\,4]{flac03}. 
The authors attributed the drop in X-ray luminosity at $\sim 3\,M_\odot$ 
to the time when stars of the ONC's age ($\sim 1$\,Myr) become fully 
radiative and, therefore, the dynamo driven X-ray activity stops. The large 
number of ONC stars with very small masses ($0.1<M/M_\odot <0.5$) has been 
undetected in X-rays.

% ----------------  Figure 5  ------------------------------------------
\begin{figure}
\epsfxsize=\columnwidth
\centering \mbox{\epsffile{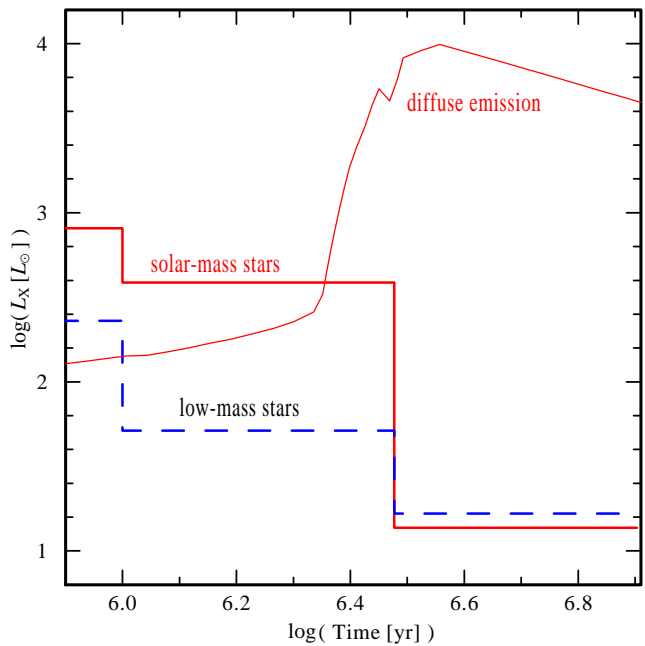}}
\caption{The collective X-ray luminosity of all cluster's low- and 
solar-mass stars (solid and dashed lines correspondingly) as a function 
of time (see text for details). For comparison the cluster wind luminosity 
is also shown (thin line).} 
\label{fig:st}
\end{figure}
%------------------------------------------------------------------------

In Table~{\ref{tab:onc}} we summarize the correlation between X-ray 
luminosity and stellar age for stars with masses ($0.5<M/M_\odot <3$)
observed in the ONC. \citet{flac03} discuss the interpretations of this 
correlations, including possible observational selection 
effects. They conclude the existence of a real age spread in the ONC. 

% ----------------  Figure 6 ------------------------------------------
\begin{figure*}
\epsfxsize=\textwidth
\centering \mbox{\epsffile{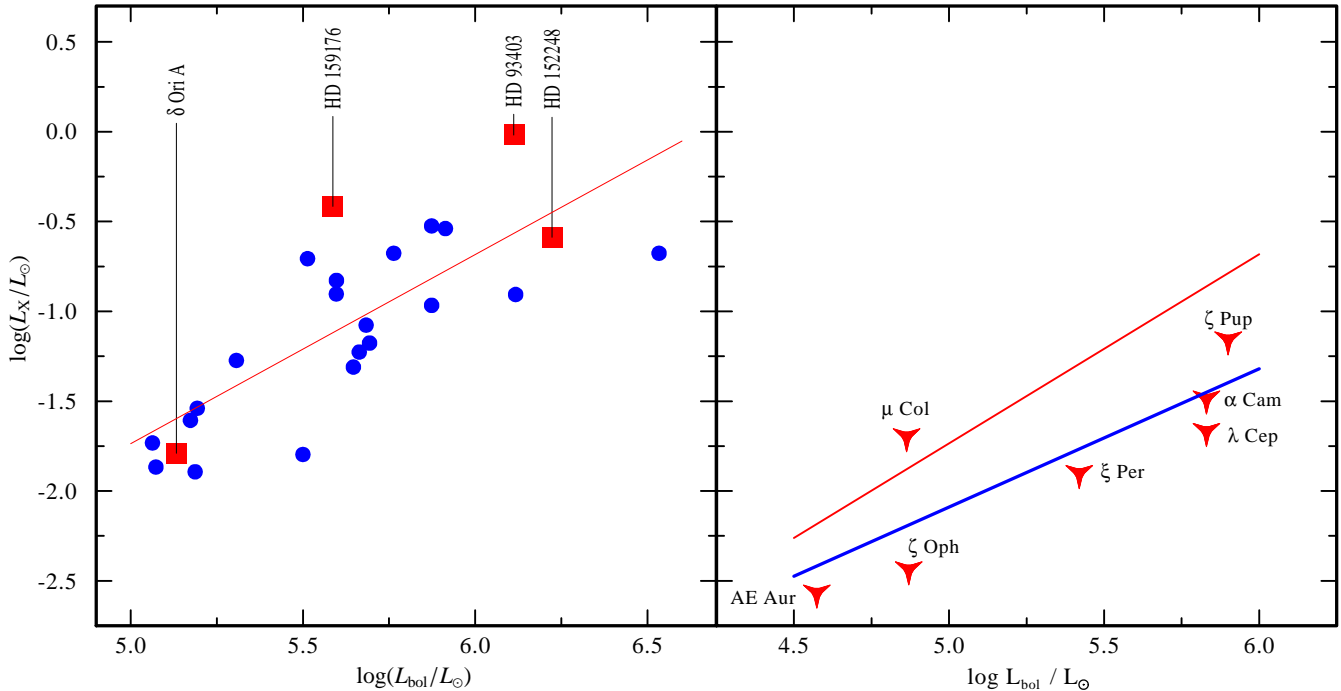}}
\caption{Bolometric luminosity of O stars as function of  their X-ray
luminosity. {\it Left panel}: Spectroscopic binaries. Circles represent the 
X-ray luminosity for the Rosat passband (0.2-2.4\,keV). Squares are for 
the XMM-Newton and Chandra data (0.2-10\,keV). The line is the linear 
regression  $\log L_{\rm X}=1.05 \log L_{\rm bol}-7.00$. {\it Right panel}: 
Single runaway stars. The thick line is the linear regression  
$\log L_{\rm X}= 0.77\log L_{\rm bol}-5.94$. The thin line is the linear 
regression obtained for the binaries.}
\label{fig:lxlbol}
\end{figure*}  
%---------------------------------------------------------------------------

Stelzer, Micela \& Neuh{\" a}user (2004) observed the central region of 
the Chamaeleon\,I South star-forming cloud with XMM-Newton. In general, 
they confirm the results obtained for the ONC and report that about 10\% 
of the low-mass stars in their sample are X-ray flaring. Most of the 
variability can be attributed to the hard band (2.5-10 keV), while the 
emission in the soft band  (0.2-2.5 keV) is more steady. Some of the 
variable stars seem to undergo irregular variability and flaring. A large 
number of stars can be detected only during short duration ($\sim 1$\,h) 
flares in the hard band.

Let us trace the evolution of the total X-ray luminosity of low- and 
solar-mass stars in our $10^6\,M_{\odot}$ idealised model 
cluster\footnote{For the estimate of the number of low-mass stars we 
assumed a lower mass cut-off of the IMF at 0.1\,$M_\odot$.}. 
The total number of stars with masses between $0.5\, ...\,1\,M_\odot$ is 
$1.7\times 10^5$, while $9.8\times10^4$ stars are in the 
$1\, ...\,3\,M_\odot$ mass bin.

Assuming that the luminosity of each star changes with age as 
specified in Table\,{\ref{tab:onc}, and adding all stars in each mass bin, 
we obtain the collective X-ray luminosity of low- and solar-mass stars 
as function of age (Fig.\,\ref{fig:st}). Individual low- and solar-mass 
stars are relatively weak X-ray sources. However, they are numerous and 
difficult to resolve spatially. Figure\,{\ref{fig:st}} demonstrates that 
in clusters younger than 3\,Myr the collective luminosity of low- and 
solar-mass stars is quite high.

To understand the contribution of low-mass stars to the total X-ray 
emission of massive clusters let us imagine how the ONC would look 
if it were located in the LMC. Numerous X-ray dim low-mass stars fill 
the whole volume of the stellar cluster, while massive stars tend to 
concentrate towards the center. Chandra's HRC observations analysed by 
\citet{flac03} covered a $30\arcmin \times 30\arcmin$ field of view, and 
were centered on the Trapezium cluster. The density of the Trapezium, 
$\approx\,5\times 10^3$ stars pc$^{-3}$ within the central 0.1\,pc, 
makes it one of the most dense clusterings known in our Galaxy. Chandra 
observations resolved 111 point sources \citep{shulz01} in the Trapezium. 
While most of these sources are faint, twelve have an X-ray luminosity 
above $10^{31}$\,erg\,s$^{-1}$ and four of those are identified with massive 
early-type stars. The ONC is located at 440\,pc. At the distance of the 
LMC, the 807 stars from Chandra's HRC observations would be all within 
$1\farcs 6\times 1\farcs 6$. The PSF of Chandra has a FWHM of $0\farcs 5$. 
Hence it would be impossible to resolve the low- and solar-mass
stellar population, while the X-ray bright stars of Trapezium would 
appear as one point source embeded in diffuse emission.  

It was found in the previous section that the cluster wind emission 
alone cannot explain the observed diffuse X-ray luminosity of NGC\,3603.  
This cluster is only about 1\,Myr old, and one may expect that stars
less massive than $1.5\,M_\odot$ are still fully convective and X-ray 
active \citep[see references in][]{flac03}. Using the standard IMF and 
the average X-ray luminosity as listed in Table\,{\ref{tab:onc}}, we 
roughly estimate the collective  X-ray luminosity of low- and solar-mass 
stars in NGC\,3603 as $L_{\rm X}^{\rm low}\sim 10^{34}$\,erg
s$^{-1}$. The sum of the cluster wind luminosity and the collective 
luminosity of low- and solar-mass stars matches the observed level of 
diffuse X-ray emission in NGC\,3603 very well, thus confirming the 
suggestion of \citet{ngc3603} and \citet{sung}. In older clusters, such as 
Quintuplet or R\,136, the low-mass stars are less active, and massive 
stars become the main sources of stellar X-ray emission.  

\subsection{ O stars}

\begin{table*}
\begin{center}
\caption{Pointing X-ray observations of O-type spectroscopic binaries and
single runaway  stars}
\smallskip
\begin{tabular}{lrrrrrrc}
\hline
\noalign{\smallskip}
Name & Exp.\,Time &Count Rate& $\log N_{\rm H}$& $d~~$ & $\log L_{\rm X}$ & 
$\log L_{\rm bol}$ & Detector \\
     & [ksec]     &[$10^2$\,ct\,s$^{-1}$]&[cm$^{-2}$]& [pc]  &[erg\,s$^{-1}$] 
&[erg\,s$^{-1}$] & \\      
\hline
\noalign{\smallskip}
\multicolumn{8}{c}{Spectroscopic binary stars} \\
\hline
\noalign{\smallskip}
HD\,57060   &3.1 &  3.55$\pm$0.4& 20.81 & 1514 & 32.15 & 39.44 & {\sc pspc}\\ 
HD\,57061   &77.7& 10.93$\pm$0.1  & 20.74 & 1514 & 32.62 & 39.46 &{\sc pspc} \\
HD\,93205   &1.4 &  5.90~~~~~~    & 21.33 & 2630 & 33.04 & 39.50 & {\sc pspc}\\
HD\,93206   &1.2 &  6.59$\pm$0~~  & 21.34 & 2512 & 33.06 & 39.46 & {\sc pspc}\\
HD\,93403   &37.8&  2.10~~~~~~    & 21.57 & 3200 & 33.57 & 39.70 &{\sc epic}\\ 
HD\,97484   &4.3 &  1.72$\pm$0.2  &20.44 & 3200 & 32.41 & 39.28 &{\sc pspc}\\  
HD\,135240  &2.5 &  7.30$\pm$0.6  & 21.18& 615  & 33.59 & 38.77 &{\sc pspc}\\
HD\,149404  &2.8 &  8.26$\pm$0.6  &21.40 & 1380 & 32.68 & 39.70 & {\sc pspc}\\ 
HD\,152218  &7.6 &  1.11$\pm$0.1  & 21.34 & 1905 & 32.05 & 38.78 &{\sc pspc}\\
HD\,152219  &7.6 &  2.21$\pm$0.2  & 20.56 & 1900 & 31.98 & 38.76 & {\sc pspc}\\
HD\,152248 &180.0&                &       & 1757 & 33.00 & 39.81 & {\sc epic}\\
HD\,152590  &8.9 &  0.91$\pm$0.1  & 20.57 & 1900 & 31.85 & 38.65 & {\sc pspc}\\
HD\,159176 &20.4&130.30$\pm$0.9 & 21.33 & 1500 & 33.17   & 39.17 &{\sc epic}\\  
HD\,165052  &3.5 & 10.41$\pm$0.6  & 21.36 & 1585 & 32.88 & 39.10 &{\sc pspc}\\
HD\,167771  &5.9 &  4.2$\pm$0~~~  & 21.40 & 1585 & 32.54 & 38.97 & {\sc pspc}\\
HD\,203064 &6.5 &  5.1$\pm$0~~~  & 20.98 & 794  & 31.79 & 39.09 & {\sc pspc}\\ 
HD\,206267  &3.1 & 10.4$\pm$0.6   & 21.49 & 780  & 32.36 & 39.25 & {\sc pspc}\\
$\delta$\,Ori\,A&2.3&176.5$\pm$2.8 &20.18&501&32.77&39.34  & {\sc acis-s} \\
$\iota$\,Ori &0.7 &158.9$\pm$4.9   & 20.30 & 501  & 32.68 & 39.18&{\sc pspc}\\
LZ\,Cep   &5.0    &  5.64$\pm$0.3  &21.33  & 583  & 31.72 & 38.66 &{\sc pspc}\\
V729\,Cyg &3.4    & 12.20$\pm$0.6  &19.94  & 1800 & 32.96 & 39.7 &{\sc pspc}\\ 
Plaskett  &3.7    & 15.60$\pm$0.7  &21.18  &1514  & 32.91 & 39.35 &{\sc pspc}\\
15\,Mon    &0.9    &  3.00$\pm$1.8  &20.35 & 692 & 32.31 & 38.89&{\sc pspc} \\ 
\hline
\noalign{\smallskip}
\multicolumn{8}{c}{Runaway single stars}\\
\hline
$\xi$\, Per  & 1.0 & 20.00$\pm$1.0 & 21.06 & 398 & 31.68 & 39.47 & {\sc pspc}\\
$\alpha$\,Cam &3.8 & 6.59$\pm$0.4 & 21.08 & 1010 & 32.10 & 39.42 & {\sc pspc}\\
$\mu$\,Col & 3.7 & 15.14$\pm$0.7 & 20.04 & 669  & 31.89 & 38.45 &{\sc pspc} \\ 
AE Aur     & 2.4 &  4.37$\pm$0.4 & 21.30 & 321  & 31.02 & 38.16 &{\sc pspc} \\
$\zeta$\,Oph & 2.0 & 39.00$\pm$2.0 & 20.78 & 154& 31.15 & 38.46 &{\sc pspc}\\  
$\lambda$\,Cep & 4.2 &  6.50$\pm$0.6 & 21.20 & 832 & 31.92 & 39.42&{\sc pspc}\\
$\zeta$\,Pup & 55.7&126.90$\pm$1.6 & 20.00 & 437& 32.43 & 39.49 & {\sc pspc}\\ 
\hline
\noalign{\smallskip}
\multicolumn{8}{l}{\small count rates are from  ``Complete Results Archive 
Sources for the {\sc pspc}`` catalog}\\
\multicolumn{8}{l}{\small unless otherwise indicated}\\
\multicolumn{8}{l}{\small HD93205:~average count rate from Rosat Catalog 
{\sc pspc} WGA Sources}\\
\multicolumn{8}{l}{\small HD93403:~$L_{\rm X}$ from \citet{ro1}}\\
\multicolumn{8}{l}{\small HD152248:~$L_{\rm X}$ from \citet{sana}}\\
\multicolumn{8}{l}{\small HD159176:~$L_{\rm X}$ from \citet{deb}}\\
\multicolumn{8}{l}{\small V729Cyg:~$L_{\rm bol}$ from \citet{v729}}\\
\multicolumn{8}{l}{\small $\delta$\,Ori:~$L_{\rm bol, X}$ as in 
\citet{deltaOri}}\\
\multicolumn{8}{l}{\small HD203064,\,$\xi$\,Per,\,$\alpha$\,Cam,\,
$\zeta$\,Oph,\,$\lambda$\,Cep,\,$\zeta$\,Pup:~$L_{\rm bol}$ 
from \citet{rep04}}
\end{tabular}
\label{tab:spb}
\end{center}
\end{table*} 
%------------------------------------------------------------------

A correlation between X-ray and bolometric luminosity for Galactic O
stars  was  suggested by \citet{sc82}. \citet{berg97} compiled a
catalog of  optically bright OB\,stars (RASSOB) based on the Rosat All
Sky Survey (RASS).  They found that $L_{\rm X}\,\approx 10^{-7}\,L_{\rm
bol}$ for the ensemble of O stars. From theoretical side, \citet{ow99}
explained this correlation for single stars invoking the
wind-momentum luminosity relation \citep{kudr00}. We decided to verify
this correlation selectively, distinguishing between binary and single
stars and using the better data available now.

\medskip
\noindent {\em Binary systems.}
We selected a sample of O+O spectroscopic binary systems 
with available Rosat pointing observations or observed by Chandra or 
XMM-Newton. In order to keep the data as uniform as possible, in case 
when observations by different instruments exist, the preference was given 
to Rosat \citep[{\em pointing} observations, in contrast to the RASS data
available to][]{berg97}. 
The data are summarized in  Table~{\ref{tab:spb}. If not 
indicated otherwise, the neutral hydrogen column density, distance and 
$L_{\rm bol}$ are taken from RASSOB. The count rates were 
retrieved from The Rosat Complete Results Archive Sources for the PSPC 
catalog, and 
transformed into  the X-ray flux assuming a Raymond-Smith thermal plasma 
with temperature $kT_{\rm X}=0.6$\,keV. For the stars with resolved X-ray 
variability, $L_{\rm X}$ was  averaged over available observations.

From linear regression analysis of our sample of spectroscopic binaries
we obtain a good correlation $\log L_{\rm X}\approx \log L_{\rm
bol} - 7$. This is rather puzzling  because  it is though that in binary
stars a significant fraction of X-rays  originates  from a colliding
winds zone. One may suggest that such a correlation is observed 
because Rosat's PSPC detector (0.2\,-\,2.4 keV) was sensitive primarily
to  the emission from individual stellar winds, which have
characteristic temperatures of about 0.6\,keV, while plasma in the
colliding wind zone might be heated to higher temperatures. If this
were true, then the stars that were observed with XMM-Newton
(0.2\,-\,10 keV) should not fit into the relation because this
instrument also detects harder radiation. The fact that there is no 
significant scatter (see left panel of Fig.\,{\ref{fig:lxlbol}})
between the binaries observed with XMM-Newton and Rosat makes us
suggest that the bulk of observed X-rays originates in the individual
stellar winds. 

\citet{deb} concluded from their study of the colliding O+O binary 
HD\,159176  that the bulk of X-rays comes from a plasma with $kT_{\rm
X}=0.6$\,keV, as expected from the hydrodynamic models of single stars,
and that the line width is consistent with the stellar wind velocity.
However, the steady-state colliding wind model systematically predicts 
too high X-ray luminosities. Nevertheless, the observed ``hard tail``
in the  X-ray spectrum of HD\,159176 most likely originates from the
colliding wind zone. \citet{deltaOri} discussed extensively the origin
of X-rays in the eclipsing system $\delta$~Ori. They as well concluded 
that the bulk of X-rays is originating in the O star wind. 

\medskip
\noindent {\em Single runaway stars.} 
If our hypothesis about stellar wind emission dominance over colliding 
wind zone emission in O binaries is correct, then there should be a
similar $L_{\rm X}\propto L_{\rm bol}$ correlation for single stars.
There is no way to select a sample of O stars which are definitely
single. However, one might select a sample of {\em  runaway stars}. The
fraction of single stars among all runaway stars is very high ($\sim
75$\%). We have chosen all single runaway stars from the O Star Speckle
Survey \citep{mason98} (see Table~{\ref{tab:spb}}) and performed a
linear regression analysis for this sample. There are only seven single
runaway stars with available Rosat pointing, therefore the result is
not statistically significant. Nevertheless, the X-ray luminosity of
single stars appears to  correlate with $L_{\rm bol}$ in a similar way
as found for the O star  binaries (Fig.~\ref{fig:lxlbol}). 

Hence we may apply the correlation $\log L_{\rm X}\approx \log L_{\rm
bol} - 7$ for O stars, disregarding whether they are single or binaries. 

{\changed 
\subsection{LBV stars}
A massive star with initial mass above $50\,M_\odot$ that began its life 
on the MS as an O type star may become very unstable near the end 
of core H-burning. Such stars are called luminous blue variables (LBVs) 
or S\,Dor variables. They populate the empirical upper luminosity boundary 
in the H-R diagram and evolve directly to the WR stage via high mass loss 
episodes.
Less luminous LBVs have initial masses less than $40\,M_\odot$ and have 
probably already been through the red supergiant stage \citep{hump}. Dozens 
of LBVs are identified in the Milky Way and the Local Group \citep{genderen}. 
Many LBVs are found in the vicinity of a young star cluster. 

Some of the LBVs are luminous X-ray sources. \citet{naze} deduce from the
Chandra observation of NGC\,346 in the SMC that the X-ray luminosity of the
binary LBV star HD\,5980 is $1.7\times 10^{34}$\,erg\,s$^{-1}$. This in fact 
exceeds the total luminosity of the  NGC\,346 cluster 
($1.5\times 10^{34}$\,erg\,s$^{-1}$). The famous Galactic LBV $\eta$\,Car 
has a similar luminosity $L_{\rm X}=1.2\times 10^{35}$\,erg\,s$^{-1}$ (Leutenegger, 
Kahn \& Ramsay 2003). \citet{pitcar} fitted the high resolution X-ray spectrum 
of $\eta$\,Car with hydrodynamic models of colliding winds. The X-ray emission 
of $\eta$~Car is similar to WR\,140 and $\gamma$~Vel: the emission 
is hard and variable, with the minimum near the phase when the star with denser 
wind is presumably in front.  

However, not all LBVs are bright in X-rays. The prototypical LBV star P\,Cyg
was observed for 97\,ksec with Rosat. Using interstellar reddening and bolometric 
luminosity from  \citet{genderen} we estimate the X-ray luminosity of P\,Cyg
as $L_{\rm X}=2\times 10^{31}$\,erg\,s$^{-1}$ resulting in 
$\log L_{\rm X}/L_{\rm bol}=-8$. Another example is the blue supergiant 
Sher\,25 associated with NGC\,3603 \citep{sher}. 
\citet{ngc3603} report that this star has no significant X-ray 
detection and attribute this to the relatively low $L_{\rm bol}$ combined 
with a low wind velocity.}

\subsection{WR stars} 

During its evolution a massive star looses a significant amount of mass 
via a stellar wind, {\changed LBV outbursts} or mass transfer in a close binary 
system,  revealing first the CNO-burning products at its surface (WN phase),  
and subsequently the He-burning products (WC phase). The mass-loss from WR 
stars 
\citep[$\dot{M}_{\rm WR}\sim 10^{-5\,..\,-6}\,M_\odot\,{\rm yr}^{-1}$ ,][]{hk98} 
is at least one order of magnitude higher than in O-type stars 
\citep[$\dot{M}_{\rm O}\sim 10^{-6\,..\,-7}\,M_\odot\,{\rm yr}^{-1}$,][]{rep04}.
The winds of WR stars are enriched by nuclear evolution products. An 
understanding  of the mechanisms by which radiative pressure drives 
WR stellar winds is currently  emerging \citep{guetz04}. 

X-ray emission from single WR stars is largely enigmatic. There are no
theoretical  models of WR winds that describe the generation of X-rays,
although the expectation  is that the same mechanism as in O winds may
operate here. This is supported by a study of instability growth rates 
in WR winds \citep{ken95}.  Meanwhile, the increasing number of 
observations of WR stars reveal a complex  picture. 
Pollock, Haberl \& Corcoran (1995) analysed the RASS observations of WR 
stars, and concluded that there  is no correlation between bolometric and 
X-ray luminosity.  

Similar to the study of O-type stars we compile the most sensitive  
up-to-date X-ray observations of WR stars (Table\,\ref{tab:wn}). We 
restrict ourselves to WN-type stars. As shown in \citet{osk03}, WC-type 
stars were not  detected in X-rays, except for a few binaries. 
Figure\,{\ref{fig:wrst}} shows the bolometric  and X-ray luminosities of 
our sample of Galactic WN stars. We first discuss the X-ray emission 
from putatively single WN stars, and then from binary stars.

%----------------  Figure 7 ------------------------------------------
\begin{figure}
\epsfxsize=\columnwidth
\centering \mbox{\epsffile{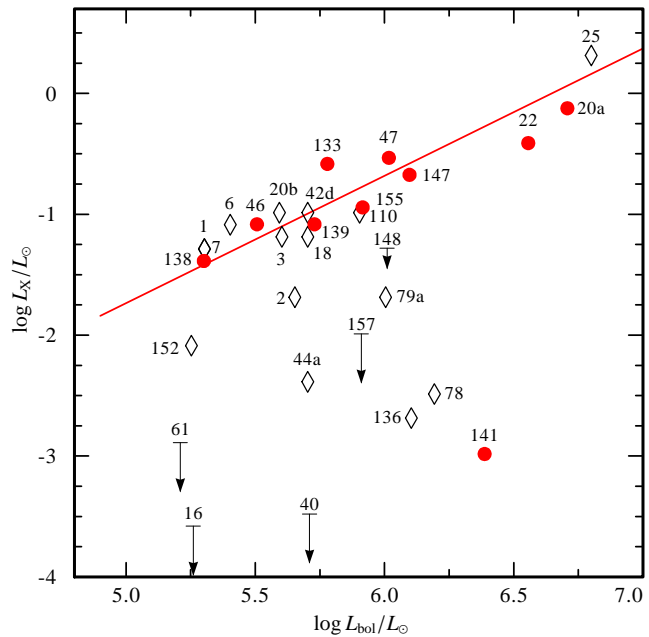}}
\caption{X-ray versus bolometric luminosity of WR stars
observed  by Chandra, XMM-Newton, or with Rosat {\sc pspc} pointings. 
Diamonds denote putatively single WN stars and circles
denote WN spectroscopic binaries. Numbers are for the stars from 
the {\sc vii}th Catalog of WR stars. Upper limits for non-detections 
are shown by the arrows. The data are from Table\,{\ref{tab:wn}}. The
thin straight line represents the correlation  between $L_{\rm X}$ 
and $L_{\rm bol}$ as found for the O-stars SBs.} 
\label{fig:wrst}
\end{figure}  
%-----------------------------------------------------------------------

\medskip
\noindent {\em Single WN stars.} Almost all bright O stars are found to
emit X-rays. Out of the 11 stars in the RASSOB sample which were 
not detected in the Rosat All Sky Survey, nine have been detected since.
Only HD\,68450  and HD\,154368 still remain undetected, 
probably because being observed only off-axis with short RASS exposures. 

In contrast, the X-ray luminosity of WN stars shows a much larger
scatter. While a couple of WN stars have significant X-ray fluxes, 
some remain below the detection limit even after long exposures. 
Correcting for their distance and reddening, which are typically
both larger than for the RASSOB stars, these WN stars still must be 
intrinsically  X-ray dim. 

%========================================================================
\begin{table*}
\caption{Sample of X-ray observations of WN-type stars}
\smallskip
\begin{center}
\begin{tabular}{rccccccc}
\hline
\noalign{\smallskip}
WR & Sp. type &$\log L_{\rm bol}$&$\log N_{\rm H}$ & d & Exp. & $\log L_{\rm X}$ & 
Detector\\
 & &[erg\,s$^{-1}$]   &[cm$^{-2}$]      &[kpc]& [ks] & [erg\,s$^{-1}$] &\\   
\hline
\noalign{\smallskip}
\multicolumn{8}{c}{Putatively single WN stars} \\
\hline
\noalign{\smallskip}
2  & WN2w & 39.24 & 21.30 & 2.51 & 9.2 & 31.9 &{\sc pspc}\\
3  & WN3  & 39.19 & 20.96 & 4.30 & 8.4 & 32.4 &{\sc epic}\\  
152& WN3w & 38.84 & 21.26 & 2.75 & 4.7 & 31.5 &{\sc pspc}\\ 
6  & WN4s & 38.99 & 20.78 & 0.97 & 9.5 & 32.5 &{\sc epic}\\   
1  & WN4s & 38.89 & 21.48 & 1.82 & 7.6 & 32.3 &{\sc epic}\\
7  & WN4  & 38.89 & 21.33 & 5.75 & 7.0 & 32.3 &{\sc pspc}\\ 
18 & WN5s & 39.29 & 21.47 & 4.57 & 7.6 & 32.4 &{\sc pspc}\\      
42d& WN5  & 39.29 & 21.85 & 4.09 & 49.3& 32.6 &{\sc acis-i}\\
44a& WN5  & 39.29 & 21.91 & 3.24 & 57.5& 31.2 &{\sc acis-s}\\ 
110& WN5-6& 39.49 & 22.02 & 1.28 &20.5 & 32.6 &{\sc epic}\\ 
20b& WN6  & 39.18 & 21.94 & 2.27 &36.2 & 32.6 &{\sc acis-i}\\
136& WN6  & 39.69 & 21.36 & 1.26 &93.9 & 30.9 &{\sc acis-s}\\ 
78 & WN7  & 39.78 & 21.28 & 1.99 & 9.9 & 31.1 &{\sc pspc}\\ 
25 & WN7a & 40.38 & 21.28 & 3.24 & 109.0& 33.9&{\sc epic}\\
79a& WN9  & 39.59 & 21.28 & 1.99 & 9.4 & 31.9 &{\sc pspc}\\
\hline
\noalign{\smallskip}
\multicolumn{8}{c}{Binary WN-type stars} \\
\hline
\noalign{\smallskip} 
46 &WN3+OB? &  39.09 &21.08& 4.07 &70.3  & 32.5 & {\sc epic}\\ 
133&WN4+O9I  &39.36 &21.15& 2.14 & 78  & 33.0 & {\sc gis+sis}\\ 
138&WN5+B?         &38.88 &21.38& 1.26 & 5.43 & 32.2 & {\sc pspc}\\
139&WN5+O6III-V&$38.9_{\rm WR}+39.1_{\rm O}$&21.50& 1.90 & 19.0 & 32.5&
{\sc pspc}\\ 
141&WN5+O5V-III &39.97 &21.66& 1.26 & ~19  & 30.6 & {\sc acis-s} \\   
47 &WN6+O5V      &39.60 &21.65& 3.80 & 151  & 33.05& {\sc epic}\\   
155&WN6+O9II&$39.08_{\rm WR}+39.29_{\rm O}$&21.40& 2.75 & 90 &32.64& 
{\sc gis+sis}\\
22 &WN7h+O9III-V &40.14 &21.02& 3.24 & 82.7     & 33.17& {\sc epic}\\ 
20a& WN6+WN6 & 39.99+39.99 &21.89& 5.75 & 32.6 & 33.46&{\sc acis-i}\\  
147&WN8(h)+B0.5V &39.68 &22.12& 0.65 & 4.9  & 32.91& {\sc hrs-i}\\
\hline
\noalign{\smallskip}
%\multicolumn{9}{c}{Binary WC-type stars} \\
%\hline
%\noalign{\smallskip} 
%48$^{1,2}$&        &       & 38.96 & 21.0 & 2.3 &    & 33.0 \\
%132$^1$   &        &       &       & 21.7 & 3.9 &    & 32.9 \\
%146$^1$   &        &       & 38.82 & 22.1 & 0.7 &    & 32.1 \\
%79$^2$    &        &       & 39.10 & 21.2 & 2.0 &    & 31.6 \\ 
%93$^1$    &        &       & 38.68 & 21.9 & 1.4 &    & 31.6 \\
%137$^1$   &        &       & 39.23 & 21.3 & 2.3 &    & 31.9 \\
%140$^{1,2,4}$&     &       & 39.4  & 21.9 & 0.3 &    & 32.9 \\
%11           &     &       &       &      &     &    &      \\
%\hline         
\multicolumn{8}{c}{Significant non-detections of WN stars} \\
\hline
\noalign{\smallskip}
61     & WN5w & 38.79 & 21.41 & 6.79 & 52.6 & $<30.7$&{\sc epic}\\
157    & WN5  & 39.49 & 21.59 & 3.39 & 5.8  & $<31.6$&{\sc pspc}\\ 
40     & WN8  & 39.29 & 21.24 & 2.26 & 16.5 & $<30.1$&{\sc epic}\\
16     & WN8h & 38.85 & 21.36 & 2.37 & 7.6  & $<30.0$&{\sc pspc}\\
124    & WN8h & 38.88 & 21.70 & 3.36 & 4    & $<32.3$&{\sc hri}\\
148    & WN8h+B3IV/BH & 39.59 & 21.62 & 8.28 & 9.9 & $<32.2$&{\sc ipc}\\ 
\hline 
\end{tabular}

\flushleft
\smallskip
\small 
X-ray luminosities are based on: 

~~{\sc pspc} count rates from ``ROSAT Complete Results 
Archive Sources for the PSPC`` catalog unless otherwise indicated

~~{\sc epic} count rates  obtained from the analysis of 
XMM-Newton archive pipeline products unless otherwise indicated

~~{\sc acis, hrs} count rates  obtained from the analysis 
of Chandra archive pipeline products unless otherwise indicated

WR\,1:~$L_{\rm X}$ from Ignace, Oskinova \& Brown (2003)

WR\,3:~Distance  from  \citet{marWR3}

WR\,6:~$L_{\rm X}$  from \citet{skinnWR6}

WR\,25:~$L_{\rm X}$  from \citet{raaWR25}

WR\,20a:~Spectral type  from \citet{rauwWR20a}.
$L_{\rm bol}$  is assigned according to the spectral type and mass

WR\,42d,\,44a,\,79a:~$L_{\rm bol}$ is assigned according 
to the spectral type

WR\,46:~$L_{\rm bol}$  from \citet{veenWR46}

WR\,79a:~Marginal detection at  12min off-axe

WR\,110:~$L_{\rm X}$   from \citet{skinnWR110}

WR\,139:~$L_{\rm bol}$  from \citet{marWR139}

WR\,141:~Located at a CCD edge

WR\,147:~Distance and extinction from \citet{pit02}

WR\,133, 155:~X-ray luminosity is roughly 
estimated from  ASCA data 
 
WR\,7, 16, 18, 124, 136, 141, 148, 157: distance, 
$E_{\rm b-v}$, and $L_{\rm bol}$ are from \citet{hk98}

WR\,22, 25, 141, 147, 155: Stellar parameters from Hamann 
et al.\ (in prep.) 

WR\,133, 138:~$L_{\rm bol}$ from Harries, Hillier \& Howarth (1998). 
For WR\,133 we add a typical luminosity of an O\,I star 
$\log{L_{\rm bol}/L_{\odot}}=5.7$
\end{center}

\label{tab:wn}
\end{table*} 
%---------------------------------------------------------------------
Most intriguing are the XMM-Newton observations of WR\,61 and WR\,40, and 
the Rosat pointing observation of WR\,16. These stars were not detected, 
and upper limits on $\log{(L_{\rm X}/L_{\rm bol})}$ are  -8.1, -9.2 and -8.9 
for WR\,61, WR\,40 and WR\,16, respectively. 

Note that WR\,40 and WR\,16 belong to the spectral subtype WN8. 
Remarkably, no WR star of this spectral type was ever detected in
X-rays (see Table\,\ref{tab:wn}). Chandra HRC-I direct imaging of the 
binary WR147~(WN8(h)+B0.5V) \citep{pit02} showed X-rays which are 
certainly  not co-spatial with the WN8 component, but associated either 
with the companion star or the colliding wind zone. There are indications 
that stellar winds from WN8 stars are  basically different from other
WN-type stars (Gr{\"a}fener,~priv.\ communication).

As can be seen from Fig.\,{\ref{fig:wrst}}, the scatter of X-ray
luminosity for the detected WN stars is extremely high, reaching up to
three orders of magnitude. Apart from the WN8 case discussed above, 
there is no correlation with the WN subtype. The outstandingly
X-ray bright star WR\,25 was carefully examined in \citet{raaWR25}, who
proposed that WR\,25 might be a binary in order to explain its rather
hard X-ray  spectrum. 

\medskip
\noindent {\em Binary WN stars.} We have examined the available X-ray
observations  of WR binary stars. One half of 22 spectroscopic binary 
WN stars listed in the {\sc vii}th Catalog of WR stars was not
detected by RASS (the detection rate is even lower  for WC-type
binaries). Among those which were detected, we select the stars with the
most sensitive  observations (Table\,{\ref{tab:wn}}). As can be seen
from Fig.\,{\ref{fig:wrst}},  the  X-ray luminosity is
proportional to the total bolometric luminosity of the system.
Moreover, the correlation is the same as found for  O+O binaries,
$\log L_{\rm X}\approx \log L_{\rm bol}-7$. The only exception is WR\,141, 
which was in Chandra's field  of view but inconveniently located at a 
CCD edge. The Chandra source detection  software recognises WR\,141.
The count rate is  $1.4\times 10^{-3}$ ct\,s$^{-1}$, yielding
$\log(L_{\rm X}/L_{\rm bol})\approx -8.7$, which is the lowest value among 
all detected massive binaries.  Therefore 
we consider this count rate being uncertain  and exclude
the star from our statistical sample to probe the $L_{\rm X}$ versus 
$L_{\rm bol}$ relation.

One should bear in mind that in general the O star is the more luminous
component in a  WR+O binary. Similar relations between $L_{\rm X}$ and 
$L_{\rm bol}$ in WN+O and O+O  binaries can be expected {\it if} the
major fraction of observed soft X-rays  originates in the individual
stellar winds rather than in the colliding winds zone. It seems that
this preposition can be confirmed observationally. \citet{Maeda} 
analysed ASCA observation of WR\,139 (V444\,Cyg). They attribute the
soft-component  emission at $kT_1\approx 0.6$\,keV and non-variable
luminosity to the individual O6 and  WN5 components of the system. The
hard component ($kT_2\approx 2$\,keV) is phase  variable and is caused
by a colliding wind shock. Our preliminary  analysis of XMM-Newton
observations of WR\,22 gives a similar result.   Overall, given the
small number of available observations, it appears  that WN+O binaries 
are quite similar to O+O binaries with respect to a  correlation
between $L_{\rm X}$ and $L_{\rm bol}$.  

Summarizing, we conclude that the major fraction of X-ray photons in 
colliding wind binaries is  emitted by the individual stellar winds of
the binary  components. The individual stellar winds typically have
thermal spectra with characteristic temperatures of  $kT_{\rm X}\approx
0.6$\,keV.  The colliding wind  zone may manifest  itself by the
presence of an additional,  somewhat  harder, either thermal or
non-thermal component.   

\medskip
\noindent {\em Binary WC stars.} {\changed  Only six WC spectroscopic binaries 
have been observed in X-rays with exposures longer than that of the RASS. ASCA  
observations of WR\,132 and WR\,113 are still waiting to be carefully 
examined. WR\,11 ($\gamma$\,Vel, WC8+O7.5III) was extensively observed 
with
Chandra and XMM-Newton \citep[e.g.][]{gamxmm}. From an analysis of the spectra
taken in different orbital phases, \citet{gamxmm} conclude that the total 
X-ray emission of the system is dominated by the material located in the wind 
collision zone. The X-ray emission from the wind of the O-type companion
is one order of magnitude smaller, and from the WC-type companion is negligible. 
For this system $\log L_{\rm X}/L_{\rm bol}=-5.9$, varying by 
$\approx 0.5$\,dex in dependence on orbital phase.  ASCA observations of 
WR\,140 (WC7+O4-5V) are reported by \citet{zheka}. This star has also been 
monitored by RXTE and Chandra. There is no detailed stellar atmosphere analysis 
available for WR\,140. We roughly estimate the bolometric luminosity of the system 
from the spectral types of the components, adopting 
$L_{\rm bol}({\rm WC7})\approx 2\times 10^5\,L_\odot$ and  
$L_{\rm bol}({\rm O6V})\approx 1.3\times 10^5\,L_\odot$. Then the ratio of
X-ray and bolometric luminosity of the system is estimated as 
$\log L_{\rm X}/L_{\rm bol}\approx -5.8$. WR\,48 ($\theta$\,Mus, WC6+O6V) 
was observed by a Rosat PSPC pointing. Assigning the same bolometric luminosity
as for WR\,140 we find $\log L_{\rm X}/L_{\rm bol}\approx -6.1$. 
WR\,79 (WC7+O5V)  was observed by XMM-Newton. We have retrieved and analysed
the archival data for this binary. The spectral types of the companions are 
similar to WR\,140 and WR\,48, therefore we use the same bolometric luminosity 
for an order of magnitude estimate, and obtain 
$\log L_{\rm X}/L_{\rm bol}\approx -7.4$. 
WR\,79 is located in the young star cluster NGC\,6231, where a large number of 
massive binaries is present \citep{garcia}. Interestingly, WR\,79 is the 
brightest X-ray source in the cluster in the 4.5-12.0\,keV band. The spectral 
energy distribution of WR\,79 is quite hard. Perhaps, by analogy with 
$\gamma$\,Vel, this hardness is due to the strong absorption that soft X-ray 
photons suffer when passing through the opaque WC wind.  

From our brief review of WC binary systems it appears that the X-ray emission 
from the colliding wind zone dominates over the emission of the individual stellar 
winds. Our tentative analysis does not reveal any $L_{\rm X}$-$L_{\rm bol}$ 
correlation for WC binaries.

Population synthesis predicts that most of the time during cluster evolution, 
the number of WC stars is much smaller than of WN stars. However in clusters 
that are about 3-4 Myr old, the number of WC stars can be comparable to the 
number of WN stars \citep{sb99}.}

Regarding to the X-ray emission from WR stars we may conclude that: (a)
the emission from binary systems does not significantly differ between
O+O and O+WN systems; (b) the level of emission from putatively single
WN stars does not correlate with  the bolometric luminosity of the
stars. Some single WR stars are quite luminous X-ray sources, while
others appear as non-emitters; {\changed (c) Although some of the WC binary 
systems can be relatively X-ray bright, this is not universal for all WC 
binaries. No X-ray emission yet detected from single WC stars.} 
Hence, it seems that $L_{\rm X}$ decreases for a significant fraction of the 
massive stars when they evolve to the WN stage, and drops even lower in the WC 
stage.

\subsection{Evolution of X-ray luminosity of massive stars}

%----------------  Figure 8 ------------------------------------------
\begin{figure}
\epsfxsize=\columnwidth
\centering \mbox{\epsffile{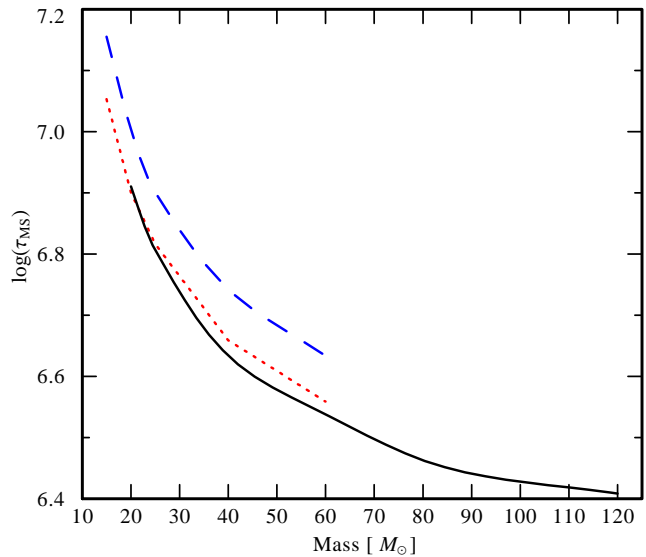}}
\caption{Lifetime in the H-burning stage, $\tau_{\rm MS}$, as function 
of the initial mass, $M_{\rm zams}$, for stars with solar metallicity.  
The results from \citet{hirschi} are for models 
with rotation ($v_{\rm rot}=300$\,km\,s$^{-1}$, dashed line) or without
rotation (dotted). Models from \citet{schaller92}, also without
rotation, extend to higher masses (solid line).} 
\label{fig:ev}
\end{figure}  
%----------------------------------------------------------------------- 

The predicted stellar lifetimes depend on the stellar evolution model. 
Recent Geneva models include the effects of  rotation 
(Hirschi, Meynet \& Maeder A., 2004). The effects of binarity on the 
evolution are included in the models by, e.g., \citet{bever}. Both effects  
extend the lifetime on the main sequence. Therefore  one
attributes a younger age to a cluster when applying these models. 
Figure\,{\ref{fig:ev}}  shows the lifetime on the H-burning stage for 
stars of different mass as given by  \citet{hirschi} and by 
\citet{schaller92}. We use the latter lifetimes in this paper. {\changed We do 
not explicitly include LBVs  because of their fast evolution 
($\sim 40\,000$\,yrs), and because little is known about their X-ray emission. 
However, an LBV star may be the dominant X-ray source in a stellar cluster, 
as observed in NGC\,346 \citep{naze}.}

% ----------------  Figure 9 ------------------------------------------
\begin{figure}
\epsfxsize=\columnwidth
\centering \mbox{\epsffile{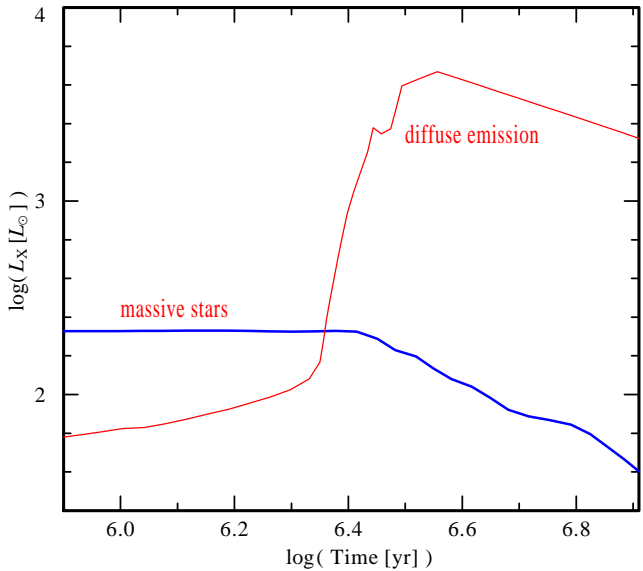}}
\caption{Evolution of the collective X-ray luminosity of massive stars
in a  cluster $M_{\rm cl}=10^6M_\odot$ as function of age.  The X-ray
luminosity of the ensemble of  massive stars strongly declines with
age.}
\label{fig:massev}
\end{figure}  
%----------------------------------------------------------------------

Figure\,{\ref{fig:massev}} shows the evolution of the collective  X-ray
luminosity  of massive stars in our model cluster. 
Considering Fig.\,\ref{fig:massev} we shall notice an interesting 
interplay between the number of massive stars and the level of diffuse
emission from the cluster wind. The cluster wind 
is fed my massive stars. If there are $N_\ast$ 
mass-loosing stars in the cluster, the cluster wind 
luminosity $L_{\rm X}^{\rm diff}\propto \dot{M}^3\propto N_\ast^3$ 
(see Eqs.\,\ref{eq:Ldif}, \ref{eq:em}) while the collective stellar
luminosity $L_{\rm X}^\ast \propto N_\ast$. Therefore, the ratio of cluster
wind to stellar emission is smaller for less massive clusters.        

Another point to address is the relation between the X-ray luminosity
of low- versus high-mass stars.  Our model predicts that  in very young
clusters the collective luminosity of low-mass stars should exceed the
collective luminosity of the high-mass stars (compare
Fig.\,\ref{fig:st} and Fig.\,\ref{fig:massev}). Nevertheless, the
massive stars are detected as point sources in, e.g., NGC\,3603, and are
not out-shone by low-mass stars. The  massive stars tend to concentrate
in the cluster core, while  low-mass stars fill the  whole volume of
the cluster. Therefore the  surface X-ray brightness due to the
low-mass stars is low, while  the X-ray bright massive stars can be
easily detected above the background. This situation lasts till the
most massive stars enter the WR phase, and start to actively feed the
cluster wind. After  $2\,..\,3$\,Myr of cluster evolution, the
emissivity of the cluster  wind exceeds the emissivity of individual
stellar X-ray sources. This  is exactly  the situation observed in the
LMC clusters, and shown in  Fig.\,\ref{fig:LMC}: the massive  stars are
seen in the youngest cluster, but sink into oblivion in the older 
clusters.        

As seen in Fig.\,\ref{fig:massev}, the X-ray luminosity declines steeply 
after $\approx 2$\,Myr of cluster evolution, when the most massive stars 
end their lives by a core collapse and a supernova explosion. Black holes 
are formed from stars with initial masses above  $\sim 21 M_{\odot}$, while 
neutron stars result from stars with lower masses \citep{woos}. 
From the analysis of the observations it appears that $\sim 2-5$\% of all 
OB stars must produce high mass X-ray binaries (HMXB) (consisting  of a 
compact object and a massive star).  This estimate is consistent with 
population synthesis studies \citep{helfand}. The lifetime of a HMXB
is limited by the nuclear time-scale of the optical companion 
($\sim 20$\,Myr).  

As can be seen from Table\,{\ref{tab:lmc}},  NGC\,2100, NGC\,1818 and
NGC\,1850 are old enough to produce compact companions in HMXBs, and are 
young enough for most non-degenerate companions still to be present. 
Adopting the quoted estimate of $\sim 2-5$\% HMXB  from all OB stars, 
one should expect 1 -- 3 HMXBs in each of these three clusters.  
However, none of them shows X-ray point sources which are bright enough 
that they could be attributed to X-ray binaries. This apparent contradiction 
may indicate a problem with the predicted HXMB formation probability.  

\section{X-ray emission from supernova remnants}

There are distinct phases in the dynamics of supernova remnants. 
Initially, the energy liberated in the collapse is deposited in the
stellar  envelope. The envelope is heated to a high temperature and
ejected with  high velocity. The expansion is uniform, with $v\propto
r$ till the mass  of swept-up material becomes significant. The highly
supersonic expansion  shock-compresses the ISM. The shocked heated ISM 
is heated and radiates by thermal  bremsstrahlung. \citet{breg} analysed 
the ROSAT observations of the complete sample of nearby supernovae that 
occurred between 1985.5 and 1994.3. They concluded, that the probability 
of an individual SN to have a luminosity higher then  $2\times
10^{39}$\,erg\,s$^{-1}$ is less then 12\%. The probability of a 
supernova to be more luminous than $6\times 10^{38}$\,erg\,s$^{-1}$ is
in  the range $9\%-51\%$. From \citet{sb99} the SN rate for a solar
metallicity  cluster is about $10^{-3}$\,SN\,yr$^{-1}$ for a $10^6
M_\odot$ cluster older than $\sim 3$\,Myr. Applying \citet{breg}
results, the rate of X-ray bright  supernovae in such a cluster can be
roughly estimated as $\sim 10^{-4}$ per  year. When an X-ray bright SN
occurs, it is likely to be the brightest  source of X-rays in its
cluster, but only for rather short time. The X-ray  brightness of SNe
declines sharp with time as $t^{-1}$ or $t^{-2}$. 

After a few\,$\times 10^2$\,years, the swept-up mass becomes greater then 
the ejected mass, and the expansion is described by the adiabatic blast-wave 
(Sedov-Taylor) solution. The radius of the remnant is given by
\begin{equation}
r_{\rm s}\,=\,0.31 (E/n_0)^{1/5}t^{2/5}~~[{\rm pc}],
\label{eq:rs}
\end{equation}  
where $E$ is the SN explosion energy in units of $10^{51}$\,erg, $n_0$ 
is the number density of the preshock ISM, and $t$ is the SNR age in 
years and it is assumed that the material has  solar composition.   
The post-shock temperature is 
\begin{equation}
kT_{\rm s}\,=\,1.8\times 10^{5}
\left(\frac{r_{\rm s}}{t}\right)^2~~[{\rm keV}].
\label{eq:ts}
\end{equation}
The frequency-integrated X-ray luminosity of the remnant is 
$L_{\rm X}^{\rm SNR}=EM\ \Lambda_{\rm X}$. The emissivity, $\Lambda_{\rm X}$,
depends on $T_{\rm s}$, electron temperature ($T_{\rm e}$), abundances, 
and the ionisation  state of the gas. For strong shock conditions 
and ideal gas behavior, the postshock-to-preshock density ratio is constant, 
$n/n_0=4$. With the dependence of the remnant radius on time in 
Eq.\,(\ref{eq:rs}), the emission measure of the remnant scales as 
$EM \propto n_0^2 t^{6/5}$.     

% ----------------  Figure 10 ------------------------------------------
\begin{figure}
\epsfxsize=\columnwidth
\centering \mbox{\epsffile{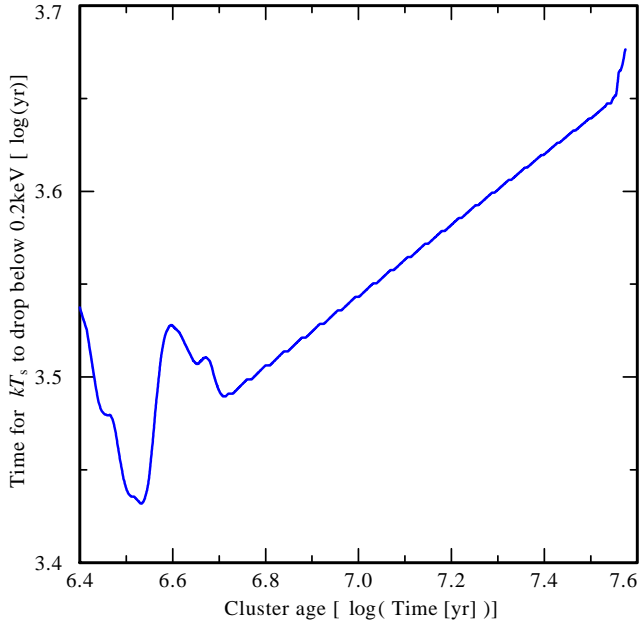}}
\caption{Time for the shock temperature $kT_{\rm s}$ of a SNR in 
Sedov-Tayler phase to drop below 0.2\,keV, as function of the model 
cluster age. 
The cluster has solar metallicity.} 
\label{fig:sedov}
\end{figure}  
%-----------------------------------------------------------------------

Borkowski, Lyerly \& Reynolds (2001) presented improved calculations of 
the X-ray spectra for
SNRs  in the Sedov-Taylor phase. Their model is implemented in the {\em
xspec}  software package. The input parameters of this model are the
shock temperature $T_{\rm s}$, electron temperature $T_{\rm e}$ and 
the ionisation timescale, determined by the product of post-shock
electron  density and the remnant's age. Using the cluster wind model
to obtain pre-shock density and $T_{\rm s}$ (see Eq.\,\ref{eq:ts}), we 
could, in principle, model SNR X-ray emission for a cluster of given age, 
and trace how the environment in the cluster affects the SNR spectra. 
However, {\sc xspec} models are not implemented over a wide enough range 
of parameters.     

Using formulae from \citet{bor} and our cluster model, we calculated 
the time it will take for a SNR in Sedov-Tayler phase to cool down below
0.2\,keV. It was assumed that $T_{\rm e}=0.5\,T_{\rm s}$ and 
$E=10^{51}$\,erg\,s$^{-1}$. The result is shown in Fig.\,\ref{fig:sedov}. 
As can be seen, the cooling time is slowly increasing for older clusters. 
However, even taking into account this small effect, in a dense cluster, 
the SNR remains a significant X-ray source, with unabsorbed 
$L_{\rm X}^{\rm SNR}\approx 10^{35\,..\,36}$\,erg\,s$^{-1}$, for only a few 
thousand years. During this time the $L_{\rm X}^{\rm SNR}$ is comparable to the 
level of the cluster wind luminosity (see Fig.\,\ref{fig:lxdif}) and the remnant 
radius is compared to the cluster core radius. However, since the cluster 
wind temperature is high and roughly constant (see lower panel in 
Fig.\,\ref{fig:lxdif}), the SNR spectrum is likely to be softer than the 
cluster wind spectrum. The supernova rate between 5 and 35 Myr in a 
$10^6 M_\odot$ cluster is about $10^{-3}$\,yr$^{-1}$ \citep{sb99}, therefore 
we may expect an enhanced level of diffuse emission for older clusters, as 
it is observed in the LMC (see Fig.\,\ref{fig:lxdif}). 
 
During the further phase in the dynamics of a SNR, as the remnant continues 
to expand, the cooling by optical emission lines become important and 
X-ray emission drops. Later the SNR expansion becomes subsonic, and the 
remnant disperses into the ISM. 
   
\section{Conclusions}

We model both the thermal X-ray emission originating in a cluster wind and 
the evolution of X-ray emission from a stellar population and compare the 
model with observations. The conclusions are: 

\noindent (1)~Effects of stellar evolution in massive clusters can explain 
the observed differences in the level of diffuse emission and numbers of 
point sources in clusters of different age.

\noindent (2)~The observed level of diffuse  X-ray emission in young massive 
star clusters is in accordance with the theory of mass-loaded cluster winds.
We retrieved and analysed X-ray observations of six prominent LMC clusters.
The striking differences in the level of  X-rays in these clusters can be 
explained by the different evolutionary stages of the clusters. The same 
holds for massive Galactic star clusters: NGC\,3603, the Arches and 
the Quintuplet.   
 
\noindent (3)~The X-ray luminosity of the cluster wind is low for the 
youngest clusters, such as NGC\,3603, where the majority of massive stars 
is still on the main sequence and, therefore, have moderate stellar winds.
  
\noindent (4)~After the most massive stars evolved to the WR stage, 
characterised by dense and fast stellar winds, the level of cluster wind 
diffuse X-ray emission rises dramatically. Powered further by supernova 
explosions, the diffuse emission of a cluster wind is nearly constant 
during about 40\,Myr. 

\noindent (5)~After about 40\,Myr of cluster evolution, the supernova rate 
is expected to drop steeply. At this stage, the cluster wind ceases.
    
\noindent (6)~The diffuse X-ray emission observed in the young Galactic
cluster NGC\,3603 cannot be  explained by a cluster wind alone. Based on
a study of the Orion  Nebular Cluster by \citet{flac03}, we model the
evolution of X-ray emission  from an ensemble of low-mass stars. We
conclude that X-ray active low-mass  stars are the dominant sources of
X-rays in clusters  younger than $\approx\,2$\,Myr. Thus, we confirm the
suggestion of \citet{ngc3603} that non-resolved low-mass stars are
responsible for  the rather high level of diffuse X-rays observed in 
NGC\,3603. 

\noindent (7)~We analyse the pointing Rosat and available XMM-Newton and 
Chandra observations of O and WN-type binaries and find that the correlation 
$L_{\rm X}=10^{-7}L_{\rm bol}$ holds for these stars.

\noindent (8)~From the study of X-ray observations of WN8 type stars we 
conclude that stars of this spectral type show no evidence of being 
X-ray sources. We speculate that the wind driving mechanism might differ 
between WN8 and other types of O and WN stars.  

\noindent (7)~ Assuming evolution with constant bolometric luminosity, 
the collective X-ray luminosity of massive stars is constant over the 
first $\sim 2$\,Myr. After the most massive stars become X-ray dim WR 
stars, the collective X-ray luminosity of massive stars declines fast.

\noindent (8)~The ratio of diffuse luminosity of the cluster wind to
the X-ray luminosity of massive stars is lower for less massive 
clusters.

\noindent (9)~If a young supernova remnant is present in a cluster, it will 
dominate the cluster X-ray emission, however only for a short time of
a few hundred years. The X-ray emission from a supernova 
remnant in the Sedov-Taylor phase is of the same order of magnitude as 
the cluster wind emission, but is expected to show a softer spectrum.  
  
\noindent (10)~Since the X-ray emission from a star cluster sensitively 
depends on its evolutionary stage, it may be used to constrain the 
cluster age and stellar population.

\section*{Acknowledgment}
This research has made use of the SIMBAD database, operated at CDS, 
Strasbourg, France, and of data obtained through the High Energy 
Astrophysics Science Archive Research Center Online Service, provided 
by the NASA/Goddard Space Flight Center. The author is grateful 
to A.\,Feldmeier for insightful discussions on hydrodynamics,  
W.-R.\,Hamann for the careful reading of the manuscript, {\changed and to 
the anonymous referee for the detailed and constructive comments}. 
The author acknowledges support from Deutsche Forschungsgemeinschaft grants 
Fe 573/1-1 and Fe 573/3-P.

{}

\label{lastpage}

\end{document}